\documentclass[prd,aps,floats,floatfix,superscriptaddress,preprintnumbers,
showpacs,eqsecnum,nofootinbib,twocolumn]{revtex4}
\usepackage{latexsym,array,theorem,mathrsfs,subfigure,bm,float}

\usepackage{psfrag}
\usepackage{amsfonts,amsmath,amssymb,latexsym,array,afterpage,
theorem,mathrsfs,bm,float,epsfig,color,graphicx,tabularx,slashbox,here,multirow}
\usepackage{indentfirst}
\newcommand{\nn}{\nonumber \\}
\newcommand{\bea}{\begin{eqnarray}}
\newcommand{\ena}{\end{eqnarray}}
\newcommand{\beann}{\begin{eqnarray*}}
\newcommand{\enann}{\end{eqnarray*}}

\newcommand{\RB}{\overset{\scriptsize  (0)}{R}}
\newcommand{\CB}{\overset{\scriptsize  (0)}{C}}
\newcommand{\Rf}{\overset{\scriptsize  (1)}{R}}
\newcommand{\GB}{\overset{\scriptsize  (0)}{G}}

\newcommand{\cGB}{\overset{\scriptsize  (0)}{{\cal G}}}

\newcommand{\gB}{\overset{\scriptsize  (0)}{g}}

\newcommand{\fB}{\overset{\scriptsize  (0)}{f}}

\newcommand{\TB}{\overset{\scriptsize  (0)}{T}}
\newcommand{\cTB}{\overset{\scriptsize  (0)}{{\cal T}}}
\newcommand{\Tf}{\overset{\scriptsize  (1)}{T}}
\newcommand{\cTf}{\overset{\scriptsize  (1)}{{\cal T}}}

\newcommand{\tauf}{\overset{\scriptsize  (1)}{\tau}}

\newcommand{\BoxB}{\overset{ \scriptsize (0)}{\Box}}
\newcommand{\nablaB}{\overset{ \scriptsize (0)}{\nabla}}

\begin{document}

\baselineskip=12pt

\title{Cosmology in ghost-free bigravity theory with twin matter fluids:
\\[,3em] The origin of ``dark matter"}
\author{Katsuki \sc{Aoki}}
\email{katsuki-a12@gravity.phys.waseda.ac.jp}
\affiliation{
Department of Physics, Waseda University,
Shinjuku, Tokyo 169-8555, Japan
}

\author{Kei-ichi \sc{Maeda}}
\email{maeda@waseda.ac.jp}
\affiliation{
Department of Physics, Waseda University,
Shinjuku, Tokyo 169-8555, Japan
}

\date{\today}

\begin{abstract}
We study dynamics of Friedmann-Lema$\hat{\i}$tre-Robertson-Walker (FLRW)
 spacetime based on the ghost-free 
bigravity theory. Assuming the coupling parameters guaranteeing 
the existence of de Sitter space as well as Minkowski spacetime, 
we find two stable attractors for spacetime with ``twin" dust matter 
fields: One is de Sitter accelerating universe and the other is 
matter dominated universe.
Although a considerable number of initial data leads to de Sitter 
universe, we also find matter dominated universe 
or spacetime with a future singularity for some initial data. 
The cosmic no-hair conjecture does not exactly hold, but 
the accelerating expansion can be found naturally. 
The $\Lambda$-CDM model is obtained as an attractor.
We also show that the dark matter component in the Friedmann equation,
which originates from another twin matter, can be  
about 5 times larger than the baryonic matter, by choosing
the appropriate coupling constants.
\end{abstract}



\maketitle

\section{Introduction}

Recent observation has confirmed the big bang scenario of 
the expanding universe. 
The cosmological parameters are determined very precisely
\cite{Planck}. 
Cosmology is now precision science.
However those observations reveal new unsolved problems in cosmology; 
dark energy and dark matter.
Dark matter could be explained by unknown elementary particles,
although there may be other possibilities.
On the other hand, dark energy, which is the origin of 
the current accelerated expansion of the Universe, is one
of the biggest mysteries in modern cosmology\cite{Perlmuter,Riess}. 
The acceleration might be due to some unknown matter
 with a strange equation of state, or
might be due to a modification of general relativity (GR). In this
paper, we are interested in the latter possibility. 
Among many modified gravity theories, 
one natural modification of GR is to
consider the possibility of a massive graviton.
The first attempt to consider a massive graviton was proposed 
by Fierz and Pauli \cite{Pauli}. 
Although a simple non-linear extension of the Fierz-Pauli massive gravity
theory  contains instabilities  called the Boulware-Deser ghost \cite{BD},
it was recently shown that the special choice of the interaction term 
can exclude  such a ghost state \cite{dRGT,gf1,gf2,gf3,gf4,gf5}\cite{deser}.
However, this theory cannot describe the flat Friedmann universe, 
if the fictitious metric for the St\"{u}ckelberg field is Minkowski's one.
One may consider an inhomogeneous metric or extend it to de Sitter metric.
If we discuss an curved fictitious geometry, it may be natural 
for it to be dynamical. 
In fact the dRGT massive gravity theory has been generalized to 
such a bigravity theory, which is still ghost-free.
It contains a massless spin-2 particle 
and a massive spin-2 particle\cite{HassanRosen}.  

Such theories with a massive graviton are also motivated by the
dark energy problem.  The accelerating universe 
may be phenomenologically described by the $\Lambda$-CDM model.
However, the theoretically expected value of a cosmological constant
(the vacuum expectation value of some fields) is 
too large to explain the observed value of dark energy.
In the massive gravity theory, the non-linear ``mass term" 
gives an effective cosmological constant.
So the graviton mass is the scale of the Hubble expansion rate,
dark energy could be explained by the massive gravity theory.
 
 Unfortunately, in the massive gravity, a flat 
Friedmann universe with a fiducial Minkowski metric 
cannot  be a solution. 
Only an open Friedmann universe solves the basic equations
\cite{MassiveCosmology3,MassiveCosmology9}. 
In order to find a flat Friedmann universe, 
the fiducial metric should 
be more generic.
One possibility is to assume  
an isotropic and inhomogeneous metric form
\cite{MassiveCosmology5,MassiveCosmology6,MassiveCosmology4,MassiveCosmology7,MassiveCosmology8}.
 However, the coupling constants are restricted in this model 
to find an accelerating universe.
The other possibility is that the fiducial spacetime is 
no longer the Minkowski's one, but it is assumed to be 
a curved spacetime. If one choose de Sitter spacetime, 
we find a flat (or closed) Friedmann universe 
as well as an open universe.
However, if both metrics are  homogeneous and isotropic, 
there appears a new type of ghost instability. 
An anisotropic fiducial metric may provide
a stable flat Friedmann universe
 \cite{Perturbation_Massive1,Perturbation_Massive2,MassiveCosmology9}.

If the fiducial spacetime is a curved spacetime, 
it may be natural for its metric to be dynamical as well.
Hence, based on a ghost-free bigravity theory, 
cosmological models are also studied 
\cite{Cosmology1,Cosmology2,Cosmology3,Cosmology4,TwinMatter2,Anisotropic1,TwinMatter1,Perturbation1,Perturbation2,Perturbation3,Perturbation4}.
In contrast to the massive gravity theory, 
these exists a self-accelerating solution for all type Freedmann
universe in the bigravity theory.

The bigravity theory includes GR with/without a
cosmological constant as a special case.
If both metric are proportional, which we call a homothetic solution,
the basic equations are reduced to
 two sets of the Einstein equations with cosmological constants,
which originate from the interaction terms of two metrics
\cite{Anisotropic2}.
When the cosmological constant is positive, we have a chance to 
find de Sitter accelerating universe
since the vacuum solution is de Sitter spacetime. 

Although a considerable literature on bigravity has discussed 
on the possibility of the accelerating universe,
it has mainly dealt with the case such that 
 the matter field interacts only with our 
physical metric. A little attention has
 been paid to the effect from an ``exotic" matter which 
interacts with another metric. Note that 
if matter fields interact with both metrics, 
it will violate the equivalence principle,
which holds in very high accuracy\cite{Will}.
Hence we have to discuss two different matter fields, 
which are decoupled each other and interact only 
through two metric interactions.
We then call them twin matter fields\cite{Bimond}.

Since GR is consistent with many experiments and observations\cite{Will}, 
homothetic solutions given by those in GR may be preferred in 
the bigravity theory as well. 
However, in homothetic solutions  
 two matters must satisfy a fine tuned condition 
\cite{Cosmology4}. 
Hence we have to include another twin matter as well as 
our matter field in the dynamics of bigravity 
and discuss whether we obtain a homothetic solution 
as an attractor.

The purpose of this study is to investigate cosmology in the ghost-free 
bigravity furthermore. 
We consider both metric are described by the FLRW metrics and 
each metric interacts with one of twin matter fields, respectively. 
We assume both are ordinary matters, which energy-momenta are conserved
 individually. However we do not assume that twin matter fields
 satisfy a fine tuned condition for 
homothetic solutions.

Since a vacuum homothetic solution can be de Sitter,
we will analyze whether this de Sitter accelerating universe 
is found as an attractor.
The result is related to the so called 
cosmic ho hair conjecture
\cite{CNHC1,CNHC2,CNHC3,CNHC4,CNHC5,CNHC6,CNHC7}, 
in which all expanding universes
 with a positive cosmological constant 
 is found  asymptotically approach the de Sitter spacetime.
If it is the case, it will guarantee naturalness of accelerating expansion 
in the bigravity theory.
The other interesting point in the present analysis is
whether another twin matter field 
behaves as dark matter.
If we find de Sitter universe with twin dark matter 
as an attractor in the present model,
the $\Lambda$-CDM cosmological model, which is 
phenomenologically favored from observations\cite{Planck},
is obtained naturally from the bigravity theory.

The paper is organized as follows. Introducing the 
ghost-free bigravity, we summarize the basic equations 
and present a homothetic solution in section \ref{HR}.
 In section \ref{cosmology}, we show the Friedmann equations 
 in bigravity theory
for the coupling constants which guarantee the existence 
of the Minkowski spacetime, and present the vacuum solutions.
 We the analyze the dynamics of the Friedmann 
equations in section \ref{dynamics}. We show that the fate of the universe is 
classified into three types (the self-accelerating  de Sitter spacetime, 
the decelerating matter dominant universe,
and the universe with a future singularity). 
The de Sitter spacetime is 
found from natural initial conditions. 
In section \ref{observation}, 
we study whether we can find an observationally consistent model in 
 our cosmological solutions. 
We find that the $\Lambda$-CDM model is obtained 
as an attractor, and dark matter can be explained by 
another twin matter for the appropriate coupling 
constants. 
We summarize our results and give some remarks in section 
\ref{summary}.

\section{Bigravity Theory}\label{HR}
\subsection{Hassan-Rosen bigravity model}
In the present papers, we will focus only on the ghost-free 
bigravity theory, although many bigravity theories have been proposed
\cite{bigravity1,bigravity2,bigravity3,bigravity4,bigravity5,bigravity6}.
 The ghost-free bigravity theory proposed by Hassan and 
Rosen \cite{HassanRosen} is described by the action
\begin{eqnarray}
S &=&\frac{1}{2 \kappa _g^2} \int d^4x \sqrt{-g}R(g)+ \frac{1}{2 \kappa _f^2}
 \int d^4x \sqrt{-f} \mathcal{R}(f) \nonumber \\
&+&
S^{[\text{m}]}(g,f, \psi_g, \psi_f)
-\frac{m^2}{ \kappa ^2} \int d^4x \sqrt{-g} \mathscr{U}(g,f) 
\,,
\label{action}
\end{eqnarray}
where $g_{\mu\nu}$ and $f_{\mu\nu}$ are two dynamical metrics, and
$R(g)$ and $\mathcal{R}(f)$ are those Ricci scalars, respectively.
  $\kappa_g^2=8\pi G$ and $\kappa_f^2=8\pi \mathcal{G}$ are 
the corresponding gravitational couplings, 
while $\kappa$ is defined by $\kappa^2=\kappa_g^2+\kappa_f^2$. 
We assume that the matter action $S^{(\text{m})}$ 
is divided into two parts:
\bea
S^{(\text{m})}(g,f, \psi_g, \psi_f)
=S_g^{[\text{m}]}(g,\psi_g)+S_f^{[\text{m}]}(f,\psi_f)
\,,
\ena
i.e.,  matter fields  $\psi_g$ and $\psi_f$ are coupled only to $g$-metric 
and to $f$-metric, respectively.
This restriction guarantees  the equivalence principle.
We call $\psi_g$ ($g$-matter) and $\psi_f$ ($f$-matter)
 twin matter fluids.

 The interaction term between two metrics is given by
\begin{equation}
\mathscr{U}(g,f)=\sum^4_{k=0}b_k\mathscr{U}_k(\gamma)
=1\,,
\end{equation}
{\setlength\arraycolsep{2pt}\begin{eqnarray}
&&\mathscr{U}_0(\gamma)=-\frac{1}{4!}\epsilon_{\mu\nu\rho\sigma} 
\epsilon^{\mu\nu\rho\sigma}\,, \nonumber \\
&&\mathscr{U}_1(\gamma)=-\frac{1}{3!}\epsilon_{\mu\nu\rho\sigma} 
\epsilon^{\alpha\nu\rho\sigma}
{\gamma ^{\mu}}_{\alpha}\,, \nonumber \\
&&\mathscr{U}_2(\gamma)=-\frac{1}{4}\epsilon_{\mu\nu\rho\sigma} 
\epsilon^{\alpha\beta\rho\sigma}
{\gamma ^{\mu}}_{\alpha}{\gamma ^{\nu}}_{\beta}\,, \\
&&\mathscr{U}_3(\gamma)=-\frac{1}{3!}\epsilon_{\mu\nu\rho\sigma} 
\epsilon^{\alpha\beta\gamma\sigma}
{\gamma ^{\mu}}_{\alpha}{\gamma ^{\nu}}_{\beta}{\gamma ^{\rho}}_{\gamma}\,, 
\nonumber \\
&&\mathscr{U}_4(\gamma)=-\frac{1}{4!}\epsilon_{\mu\nu\rho\sigma} 
\epsilon^{\alpha\beta\gamma\delta}
{\gamma ^{\mu}}_{\alpha}{\gamma ^{\nu}}_{\beta}{\gamma ^{\rho}}_{\gamma}
{\gamma ^{\sigma}}_{\delta}\,,
\nonumber
\end{eqnarray}}
where $b_k$ are coupling constants, while ${\gamma^{\mu}}_{\nu}$ is 
defined by 
\begin{equation}
{\gamma^{\mu}}_{\rho}{\gamma^{\rho}}_{\nu}
=g^{\mu\rho}f_{\rho\nu}
\,. 
\label{gamma2_metric}
\end{equation}
In order to take the square root to obtain the explicit form of
 ${\gamma^{\mu}}_{\nu}$, we shall introduce the tetrad systems,
$\{e_\mu^{(a)}\}$ and $\{\omega_\mu^{(a)}\}$, 
which are defined by
\begin{equation}
g_{\mu\nu}=\eta_{ab}e_\mu^{(a)}e_\nu^{(b)}
\,,~~
f_{\mu\nu}=\eta_{ab}\omega_\mu^{(a)}\omega_\nu^{(b)}
\,,
\end{equation}
with an additional constraint 
$e^\mu{}_{(a)} \omega_{\mu (b)}
=e^\mu{}_{(b)} \omega_{\mu (a)}$.
This constraint guarantees that the tetrad description is 
equivalent to the metric description.

We then find 
\bea
{\gamma^{\mu}}_{\nu}=\epsilon \eta_{ab} e^{\mu}{}^{(a)}\omega_\nu^{(b)}
\,,
\ena
where $\epsilon=\pm 1$ comes from the square root. 
As for the directions of tetrads, we 
choose that $e_\mu^{(0)}dx^\mu$ and $\omega_\mu^{(0)}dx^\mu$ 
are future-directed for $dt>0$.
Changing the sign of $\epsilon$ corresponds to 
 the following transformation
\begin{equation}
{\gamma^{\mu}}_{\nu} \leftrightarrow  -{\gamma^{\mu}}_{\nu}
\,,
\end{equation}
for which the interaction term is invariant 
by changing the sign of 
the coupling constants as
\bea
b_k\leftrightarrow (-1)^k b_k ~~~(k=0-4)
\,.
\ena
Note that $\mathscr{U}_0(\gamma)$  and 
 $\mathscr{U}_4(\gamma)$ do not contribute to the 
equations of motion for $f_{\mu\nu}$ and $g_{\mu\nu}$, respectively, 
because
\begin{eqnarray}
&\sqrt{-g}\mathscr{U}_0(\gamma)=\sqrt{-g}, \\
&\sqrt{-g}\mathscr{U}_4(\gamma)=\sqrt{-f}.
\end{eqnarray}
which are just cosmological constants in $g$-spacetime
 and $f$-spacetime\cite{footnote}, respectively. 
The interaction term is also written by another tensor  
defined by
$\mathcal{K}^{\mu}_{\nu}=\delta^{\mu}_{\nu}-{\gamma^{\mu}}_{\nu}$
as 
\begin{equation}
\mathscr{U}(g,f)=\sum^4_{k=0}c_k\mathscr{U}_k(\mathcal{K}),
\end{equation}
where the relations between $\{b_k\}$ and $\{c_k\}$ are
{\setlength\arraycolsep{2pt}\begin{eqnarray}
c_0&=&b_0+4b_1+6b_2+4b_3+b_4, \nonumber \\
c_1&=&-(b_1+3b_2+3b_3+b_4), \nonumber \\
c_2&=&b_2+2b_3+b_4, \\
c_3&=&-(b_3+b_4), \nonumber \\
c_4&=&b_4. \nonumber
\end{eqnarray}}

If we require a flat Minkowski spacetime to be a solution of the 
field equations, we have to impose the following condition
\begin{equation}
c_0=c_1=0
\,.
\label{c_0=0}
\end{equation}
Since $m$ is assumed to be 
the mass of graviton in the Minkowski background 
 in massive gravity limit, 
we should set  
\begin{equation}
c_2=-1
\,.\label{c_2=-1}
\end{equation}
This quadratic term $\mathscr{U}_2(\mathcal{K})$ 
gives the Fierz-Pauli term in the limit of linear 
massive gravity theory
\cite{Pauli}.

As a result, $\{b_k\}$ are also given by 
two free coupling constants $c_3$ and $c_4$ as
{\setlength\arraycolsep{2pt}\begin{eqnarray}
b_0&=&4c_3+c_4-6, \nonumber \\
b_1&=&3-3c_3-c_4, \nonumber \\
b_2&=&2c_3+c_4-1, \label{flatsol}\\
b_3&=&-(c_3+c_4), \nonumber \\
b_4&=&c_4. \label{coupling_Minkowski}
\end{eqnarray}}

In the present paper, we shall focus on this choice 
of the coupling constants except for some special case
such as the partially massless theory.

In de Sitter spacetime, it is known that
 there exists the so-called Higuchi bound
\begin{equation}
m_{FP}^2=\frac{2}{3}\Lambda\,,
\end{equation}
where $m_{FP}$ is the mass of spin-2 particle in the linear massive gravity 
theory 
\cite{Pauli} and $\Lambda$ is a cosmological constant. 
Beyond this bound, no ghost appears 
 and then five modes of the massive graviton can propagate properly, while 
below the bound, the helicity-zero mode
 becomes a ghost \cite{Higuchi1,Higuchi2}. 
At the exact bound value, however,
the helicity-zero mode is decoupled and a new gauge symmetry 
appears. Such a theory is often referred to partially massless (PM).
 A non-linear extension is known as the PM massive gravity, and
the extension to bigravity theory (the PM bimetric theory) is also discussed
\cite{PM0,PM}.
These PM theories are characterized by the coupling constants such that
\begin{equation}
b_1=b_3=0
\,,~~
b_0=3b_2\,{\kappa_f^2}/{\kappa_g^2}
\,,~~
b_4=3b_2\,{\kappa_g^2}/{\kappa_f^2}
\,. 
\label{PM}
\end{equation} 
We shall discuss this special case separately in Appendix.

\subsection{The equations of motion}
Taking the variation of the action with respect to $g_{\mu\nu}$ and
$f_{\mu\nu}$, we find two sets of the Einstein equations:
{\setlength\arraycolsep{2pt}\begin{eqnarray}
{G ^{\mu}}_{\nu} &=&
\kappa _g^2 ( {T ^{ [\gamma ] \mu} }_{\nu} 
+ {T^{\text{[m]} \mu} }_{\nu} ) \label{g-equation}, \\
{ \mathcal{G} ^{\mu}}_{\nu} &=& \kappa _f^2
( {\mathcal{T} ^{ [\gamma ] \mu} }_{\nu} 
+ {\mathcal{T}^{\text{[m]} \mu} }_{\nu} ), \label{f-equation}
\end{eqnarray}}
where ${G ^{\mu}}_{\nu}$ and ${ \mathcal{G} ^{\mu}}_{\nu} $ are the Einstein 
tensors for $g_{\mu\nu}$ and $f_{\mu\nu}$, respectively. 
The matter energy-momentum tensors 
are given by 
\bea
{ T^{\text{[m]}}}_{\mu\nu} &=&-2{\delta S_g^{[\text{m}]}
\over \delta g^{\mu\nu}}
\nn
{\mathcal{T}^{\text{[m]}}}_{\mu\nu}&=&-2{\delta S_f^{[\text{m}]}
\over \delta f^{\mu\nu}} 
\,.
\ena
 The energy-momentum tensors of ``gravitons", which are from the interaction 
term, are given by
{\setlength\arraycolsep{2pt}\begin{eqnarray}
&&{T ^{ [\gamma ] \mu} }_{\nu}=\frac{m^2}{\kappa^2} \
 ({\tau ^{\mu}}_{\nu} - \mathscr{U} {\delta ^{\mu}}_{\nu} \label{T(g)} ), 
\\
&& {\mathcal{T} ^{ [\gamma ] \mu} }_{\nu}  = -\frac{\sqrt{-g}}{\sqrt{-f}} 
\frac{m^2}{\kappa^2} {\tau ^{\mu}}_{\nu} \label{T(f)},
\end{eqnarray}}
with
{\setlength\arraycolsep{2pt}\begin{eqnarray*}
\tau^\mu_{~\nu}&=&
\{b_1\,\mathscr{U}_0+b_2\,\mathscr{U}_1+b_3\,\mathscr{U}_2
+b_4\,\mathscr{U}_3\}\gamma^\mu_{~\nu} \notag 
\nonumber \\
&-&\{b_2\,\mathscr{U}_0+b_3\,\mathscr{U}_1+b_4\,
\mathscr{U}_2\}(\gamma^2)^\mu_{~\nu}  \\
&+&\{b_3\,\mathscr{U}_0+b_4\,\mathscr{U}_1\}(\gamma^3)^\mu_{~\nu} 
\notag  \\
&-&b_4\,\mathscr{U}_0\,(\gamma^4)^\mu_{~\nu}
\,.
\end{eqnarray*}}

The energy-momenta of matter fields are assumed to be 
conserved individually as
\begin{equation}
\overset{(g)}{\nabla} _{\mu}{T^{ [\text{m}] \mu} }_{\nu}=0\,,\;
\overset{(f)}{\nabla} _{\mu} {\mathcal{T} ^{ [\text{m} ] \mu} }_{\nu} =0
\,, 
\label{c1}
\end{equation}  
where $\overset{(g)}{\nabla} _{\mu}$ and $\overset{(f)}{\nabla} _{\mu}$ are 
covariant derivatives with respect to $g_{\mu\nu}$ and $f_{\mu\nu}$. 
From the contracted Bianchi identities for 
\eqref{g-equation} and \eqref{f-equation}, 
the conservation of energy-momenta for ``gravitons" are 
also guaranteed as
\begin{equation}
\overset{(g)}{\nabla} _{\mu}{T ^{ [\gamma] \mu} }_{\nu}=0\,,\;
\overset{(f)}{\nabla} _{\mu} {\mathcal{T} ^{ [\gamma] \mu} }_{\nu} =0
\,.
\label{c2}
\end{equation}

\subsection{Homothetic solution}
\label{homothetic}
Before going to discuss cosmology, first we give one simple solution,
in which we assume that two metrics are proportional;
\bea
f_{\mu\nu}=K^2\, g_{\mu\nu}
\,,
\ena
where $K$ is a scalar function.
In this case, since we find the tensor $\gamma^\mu{}_\nu=K\,
\delta^\mu{}_\nu$, the energy-momentum tensors from the 
interaction term is given by
\begin{eqnarray*}
\kappa_g^2{T ^{ [\gamma ] \mu} }_{\nu} &=&-\Lambda_g(K)\delta^\mu{}_\nu
\,,
\\
\kappa_f^2{\mathcal{T} ^{ [\gamma ] \mu} }_{\nu} &=&-\Lambda_f(K)
\delta^\mu{}_\nu
\,,
\end{eqnarray*}
 where 
\begin{eqnarray*}
\Lambda_g(K)&=&m_g^2\,\left(b_0+3b_1 K+3b_2 K^2+b_3 K^3\right)
\\
\Lambda_f(K)&=&m_f^2\,\left(b_1+3b_2 K^{-1}+3b_3 K^{-2}+b_4K^{-3} \right)
\,,
\end{eqnarray*}
with 
\bea
&&
m_g^2={\kappa_g^2\over \kappa^2}\,m^2
\,,~~{\rm and}~~~
m_f^2={\kappa_f^2\over \kappa^2}
\,m^2
\nn
&&
\hskip .5cm (m_g^2+m_f^2=m^2).
\ena

From the energy-momentum conservation (\ref{c2}), 
we find that $K$ is a constant.
As a result, we find two sets of the Einstein
equations with cosmological constants
$\Lambda_g$ and $\Lambda_f$:
\bea
G_{\mu\nu}(g)+\Lambda_g\,g_{\mu\nu}&=&\kappa_g^2 {T^{\text{[m]}}}_{\mu\nu}
\label{homothetic_g}
\\
\mathcal{G}_{\mu\nu}(f)+\Lambda_f\,f_{\mu\nu} &=& \kappa _f^2
 {\mathcal{T}^{\text{[m]} } }_{\mu\nu} 
\label{homothetic_f}
\,.
\ena
Since two metrics are proportional, we have the constraints on 
the cosmological constants and matter fields as
\bea
&&\Lambda_g(K)=K^2\Lambda_f(K)
\label{eq_K}
\\
&&{\mathcal{T}^{\text{[m]} } }_{\mu\nu} 
=K^2\,{T^{\text{[m]}}}_{\mu\nu}
\,.
\ena
Since (\ref{eq_K}) is a quartic equation of $K$, 
we have at most four real roots of $K$, which give 
four different cosmological constants.
The basic equations  (\ref{homothetic_g}) 
(or (\ref{homothetic_f})) are just the Einstein equations 
in GR with a cosmological constant.
Hence any solutions in GR with a cosmological constant are always
the solutions in the present bigravity theory.
We shall call these solutions homothetic solutions because of
the proportionality of two metrics.

If we assume a flat Minkowski spacetime is one of the solutions
in the case of vacuum state,
that is, if the coupling constants are given by (\ref{c_0=0})
and (\ref{c_2=-1}), or  (\ref{coupling_Minkowski}),
we find $K=1$ is always one of the solutions, which gives zero cosmological 
constant ($\Lambda_g(1)=\Lambda_f(1)=0$).
Among the rest three solutions of $K$, if we find a positive 
cosmological constant, we obtain 
an accelerating expansion of the universe, which evolves into  
de Sitter solution, if the cosmic ho hair conjecture holds.

In Appendix, we present the perturbation equations for 
homothetic solutions. The mass of massive mode in the homothetic background 
is given by 
\bea
m_{\rm eff}^2=\left(m_g^2+{m_f^2\over K^2}\right)
\left(b_1 K+2b_2K^2+b_3K^3\right)
\,.
\label{graviton_mass}
\ena
The homothetic de Sitter solution is 
stable against linear perturbations.

\section{Cosmology in bigravity}
\label{cosmology}
Based on the ghost-free bigravity theory as well as the massive gravity 
theory, many authors have studied cosmological models.
In this paper, we analyze the details of the evolution of the universe 
including both matter fields
and study whether an accelerating expansion is naturally found in the late
 time. This is related to the so-called cosmic no hair conjecture in general 
relativity (GR), in which 
de Sitter solution is an attractor for generic initial conditions 
if there exists a cosmological constant. 
Especially, we focus on the effect of matter fields including 
$f$-matter, which has not been studied so much.

\subsection{FLRW universe}

Now we discuss the FLRW spacetime, which metrics 
are given by\cite{footnote1}
{\setlength\arraycolsep{2pt}\begin{eqnarray}
ds_g^2&=&-N_g^2(t)dt^2+a_g^2(t) \left( \frac{dr^2}{1-kr^2} +r^2 d\Omega ^2
\right)\,, \label{g_munu} \\
ds_f^2&=&-N_f^2(t)dt^2+a_f^2(t) \left( \frac{dr^2}{1-kr^2} +r^2 d\Omega ^2
\right)\,, \label{f_munu}
\end{eqnarray}}
where $N_g$ and $N_f$ are lapse functions, while 
$a_g$ and $a_f$ are scale factors for $g_{\mu\nu}$ and 
$f_{\mu\nu}$, respectively.
Since those variables must be positive, 
we choose the tetrads as
\bea
\{e^{(a)}_\mu \}&=&{\rm diag} \left(N_g, {a_g\over \sqrt{1-kr^2}},
 a_g, a_g\sin\theta\right)
\label{tetrad_g}
\\
\{\omega^{(a)}_\mu \}&=&{\rm diag} \left(N_f, {a_f\over \sqrt{1-kr^2}},
 a_f, a_f\sin\theta\right)
\label{tetrad_f}
\ena

Hence the interaction tensor is given by
\bea
\gamma^\mu_{~\nu}=\epsilon\, {\rm diag}\left(A, B, B, B\right)
\,,
\ena
where $A=N_f/N_g$ and $B=a_f/a_g$.

The cosmic times for $g$- and $f$-metrics are defined by
\begin{equation}
\tau_g =\int N_g(t)dt\,,~~\tau_f =\int N_f(t)dt
\,,
\end{equation}
respectively. 
Using the gauge freedom, in what follows, 
we set $N_g=1$, in which gauge choice, 
the time coordinate $t$ 
is the same as the cosmic time of $g$-metric.

Setting $\tilde{A}=\epsilon A, \tilde{B}=\epsilon B$, 
we find that the interaction energy-momentum tensors are given by
{\setlength\arraycolsep{2pt}\begin{eqnarray}
{T_g ^{ [\gamma ] \mu} }_{\nu}&=&\text{diag}
\left[-\rho^{[\gamma]}_g,P^{[\gamma]}_g,P^{[\gamma]}_g,P^{[\gamma]}_g\right],
 \\
{T_f ^{ [\gamma ] \mu} }_{\nu}&=&\text{diag}
\left[-\rho^{[\gamma]}_f,P^{[\gamma]}_f,P^{[\gamma]}_f,P^{[\gamma]}_f\right]
\end{eqnarray}}
where
{\setlength\arraycolsep{2pt}\begin{eqnarray}
\rho^{[\gamma]}_g&=&\frac{m^2}{\kappa^2}(b_0+3  
b_1\tilde{B}+3b_2\tilde{B}^2+b_3  \tilde{B}^3),\\
P^{[\gamma]}_g&=&-\frac{m^2}{\kappa^2}\Big[
b_0+ b_1(\tilde{A}+2\tilde{B})
\nn
&&
~~+b_2(2\tilde{A}\tilde{B}
+\tilde{B}^2)+  b_3\tilde{A}\tilde{B}^2\Big]\,,\\
\rho^{[\gamma]}_f&=&\frac{m^2}{\kappa^2}
\left(b_4+\frac{3 b_3}{\tilde{B}}
+\frac{3b_2}{\tilde{B}^2}+\frac{ b_1}{\tilde{B}^3}\right),\\
P^{[\gamma]}_f&=&-\frac{m^2}{\kappa^2}\Big[ b_4
+ b_3\left(\frac{1}{\tilde{A}}
+\frac{2}{\tilde{B}}\right) 
\nn
&&
~~+b_2\left( \frac{2}{\tilde{A}\tilde{B}}+\frac{1}{\tilde{B}^2}\right)
+\frac{ b_1}{\tilde{A}\tilde{B}^2}\Big]\,.
\end{eqnarray}}

We assume that  twin matter fields ($g$-matter and $f$-matter fluids)
are described by perfect fluids:
\begin{eqnarray*}
&&{T_g ^{ [\text{m}] \mu} }_{\nu}=\text{diag}
\left[-\rho_g(t), P_g(t), P_g(t), P_g(t)\right]\,, \\
&&{T_f ^{ [\text{m}] \mu} }_{\nu}=\text{diag}
\left[-\rho_f(t), P_f(t), P_f(t), P_f(t)\right]\,.
\end{eqnarray*}
Assume that the universe consists of dust (non-relativistic matter) 
and radiation (relativistic matter) for 
twin matter fluids. 
From the conservation equations,
\begin{eqnarray}
&&
\dot{\rho}_g+3\frac{\dot{a}_g}{a_g}(\rho_g+P_g)=0\,,
\nn
&&
\dot{\rho}_f
+3\frac{\dot{a}_f}{a_f}(\rho_f+P_f)=0\,,
\end{eqnarray}
where the dot denotes the derivative with respect to $t$,
 the energy densities are described by the scale factors as 
\bea
\kappa_g^2\rho_g&=&\kappa_g^2(\rho_{g,{\rm m}}+\rho_{g,{\rm r}})=
{c_{g,{\rm m}}\over a_g^3}+{c_{g,{\rm r}}\over a_g^4}
\nn
\kappa_f^2\rho_f&=&\kappa_f^2(\rho_{f,{\rm m}}+\rho_{f,{\rm r}})=
{c_{f,{\rm m}}\over a_f^3}+{c_{f,{\rm r}}\over a_f^4}
\label{energy_density_matter}
\,,
\ena 
where $c_{g,{\rm m}},c_{g,{\rm r}},c_{f,{\rm m}}$ and $c_{f,{\rm r}}$ 
are positive integration constants.

The Einstein equations with the metric ansatz 
(\ref{g_munu}) and (\ref{f_munu}) 
are reduced to the Friedmann equations:
{\setlength\arraycolsep{2pt}\begin{eqnarray}
{H_g^2} +\frac{k}{a_g^2}&=&
{\kappa_g^2\over 3}\left[\rho^{[\gamma]}_g +\rho_g\right]\,, 
\label{g1} \\
{H_f^2} +\frac{k}{a_f^2} &=&
{\kappa_f^2\over 3}\left[\rho^{[\gamma]}_f +\rho_f\right]\,,  
\label{f1}
\end{eqnarray}}
where 
\bea
H_g={\dot a_g\over  a_g}\,,~~H_f={\dot a_f\over N_f a_f}
\ena
are the Hubble expansion parameters.
 
The conservation equations for 
${T_g ^{ [\gamma ] \mu} }_{\nu}$ and  ${T_f ^{ [\gamma ] \mu} }_{\nu}$
are reduced to one equation:
\begin{equation}
\left(\frac{\dot{a_f}}{\dot{a_g}}-A \right) 
(b_1+2b_2\tilde{B}+ b_3\tilde{B}^2)=0.
\end{equation}
These are two cases: The first parentheses vanishes or the second one does so. 
If the second parentheses vanishes,
 $\tilde{B}$ is a constant, and then 
$\rho^{[\gamma]}_g(\tilde{B})$ and $\rho^{[\gamma]}_f(\tilde{B})$ 
are also constant. As a result, the Friedmann equations 
\eqref{g1} and \eqref{f1} are the same as the ordinary ones in GR 
with a cosmological constant. 
Since the evolution of the universe is well analyzed in GR,
we will not discuss this case furthermore.

Thus, we assume that the first parentheses vanishes. 
This condition holds when
\begin{equation}
H_g=B H_f
\,.
\label{Hubble_relation}
\end{equation}

From two Friedmann equations (\ref{g1}), (\ref{f1}) 
with the condition (\ref{Hubble_relation}),
we find 
one algebraic equation 
\begin{equation}
\kappa_g^2 \left[ {\rho}^{[\gamma]}_g(\tilde{B})+{\rho}_g(a_g)
\right]
-{\kappa_f}^2 \tilde{B}^2 \left[{\rho}^{[\gamma]}_f(\tilde{B})
+ {\rho}_f(a_f)\right]
=0
\,. 
\label{A2} 
\end{equation}
Since $a_f=B a_g=\epsilon \tilde{B}a_g$, 
this equation gives the relation between 
$\tilde{B}$ and $a_g$. 
It also provides us some information about 
the interaction term ${\rho}^{[\gamma]}_g$ and 
${\rho}^{[\gamma]}_f$ in terms of twin matter 
fluids, which will be used in the discussion about 
dark matter later.

The above equation (\ref{A2}) with (\ref{energy_density_matter})
is rewritten into a quartic equation 
for $a_g$ as
\begin{equation}
\tilde{B} C_{\Lambda}(\tilde{B})a_g^4
+\tilde{B}C_{\text{m}}(\tilde{B})a_g+C_{\text{r}}(\tilde{B})
=0, 
\label{A2'}
\end{equation}
where
{\setlength\arraycolsep{2pt}\begin{eqnarray}
C_{\Lambda}(\tilde{B})&=&\tilde{B}\left[\kappa_g^2
 {\rho}^{[\gamma]}_g(\tilde{B})
-{\kappa_f}^2 \tilde{B}^2 {\rho}^{[\gamma]}_f(\tilde{B})\right]
\\
&=&\kappa_g^2\tilde{B} \left(b_3\tilde{B}^3+3b_2 \tilde{B}^2
+3  b_1 \tilde{B}+b_0
\right)\nonumber \\
&&
-\kappa_f^2\left(b_4 \tilde{B}^3
+3 b_3 \tilde{B}^2 +3 b_2\tilde{B}+ b_1  
\right), \nonumber \\
C_{\text{m}}(\tilde{B})&=& 
c_{g,\text{m}}\tilde{B}-\epsilon  c_{f,\text{m}}
\,,
\\
C_{\text{r}}(\tilde{B})&=&c_{g,\text{r}}\tilde{B}^2
-c_{f,\text{r}}\,.
\end{eqnarray}}

Solving (\ref{A2'}), we obtain the relation 
$a_g=a_g(\tilde{B})$ and then $a_f=\epsilon\tilde{B} a_g(\tilde{B})$.
Plugging this relation into the Friedmann equation (\ref{g1}),
we find the equation for $\tilde{B}$ as
\bea
\left({d\tilde{B}\over dt}\right)^2+V_g(\tilde{B})=0
\label{eq_B}
\,,
\ena 
where the potential for $\tilde B$ is given by 
\begin{eqnarray*}
V_g(\tilde{B})={a_g^2\over a'^2 _g}\left[
{k\over a_g^2(\tilde{B})}-{1\over 3}\left(
\kappa_g^2\rho^{[\gamma]}_g(\tilde{B})
+{c_{g,{\rm m}}\over a_g^3(\tilde{B})}+{c_{g,{\rm r}}\over a_g^4(\tilde{B})}
\right)\right]
\end{eqnarray*}
with
\begin{eqnarray*}
a'_g=
-{\left(C_{\Lambda}+\tilde{B}C_{\Lambda}'\right)a_g^4+
(2c_{g,{\rm m}}\tilde{B}-\epsilon c_{f,{\rm m}})a_g+2c_{g,{\rm r}}\tilde{B}
\over \tilde{B}(4C_{\Lambda}a_g^3+C_{\rm m})}
\,.
\end{eqnarray*}
A prime denotes the derivative with respect to $\tilde{B}$.

\subsection{Vacuum solutions}
Since the matter energy densities drop as the universe expands,
we may expect the spacetime evolves into 
a vacuum state asymptotically unless 
the spacetime encounters a singularity.

So before solving the dynamical equation (\ref{eq_B}),
we first analyze the vacuum solutions. 
Eq. (\ref{A2'})
is now simple as  
\bea
C_\Lambda(\tilde{B})=0
\label{eq_B0}
\,.
\ena
This equation is the same as (\ref{eq_K}).
Since $\tilde{B}$ is a constant, we find $A=B=K$, which gives 
the homothetic solution discussed in \S. \ref{homothetic}.

We have at most four real roots $\tilde{B}_\ell~(\ell=1,\cdots,4)$ for
Eq. (\ref{eq_B0}).
If we assume the existence of Minkowski space, 
we always have one trivial root $\tilde{B}_{\rm (M)}=1
~(B_{\rm (M)}=1, \epsilon=1)$, which 
gives zero cosmological constant.
The rest three  roots can be all real
 or one real and two complex.

First we shall look for de Sitter solution. 
When the following conditions are satisfied, we find 
one de Sitter solution, which is given by 
$B_{\rm (dS)}$ with $\epsilon=1$:
\begin{eqnarray*}
&c_3<0\,,~c_3+c_4<0\,,~2c_3^2+3c_4>0&\, \;[\text{region (1)}] \,, 
\label{b4}
\nn
&c_3>3\,,~3c_3+c_4<3\,,~2c_3^2+3c_4>0&\, \;[\text{region (2)}] \,. 
\label{b0}
\end{eqnarray*}
We show the typical examples (Models A and B) 
for these regions in Table \ref{epsilon1}:
\begin{table}[H]
\begin{center}
  \begin{tabular}{|c|c||c|c|c|r|c|}
\hline 
Model&$(c_3,c_4$)&region&$\epsilon$&$B_\ell$&$\Lambda_g$~~~~~~~&vacuum \\ 
\hline 
\hline 
A& $(-1, 0)$&(1)&$-1$&$0.523476$&$-22.0323m_g^2$&AdS$_1$ 
\\ \cline{4-7} 
&&&1&1&0&M  \\ \cline{4-7} 
&&&1&1.67319&$-0.394464m_g^2$&AdS$_2$ \\ \cline{4-7} 
&&&1&$6.85028$&$12.4267m_g^2$&dS \\ 
 \hline \hline 
B& ($4, -10)$&(2)&$-1$&$1.91031$&$-80.4017 m_g^2$&AdS$_1$  \\ \cline{4-7} 
&&&$1$&$1$&$0$&M \\ \cline{4-7}  
&&&$1$&$0.145979$&$0.264813m_g^2$&dS \\ \cline{4-7} 
&&&1&$0.59766$&$-0.140902m_g^2$&AdS$_2$   \\ 
\hline 
 \end{tabular}
    \caption{In the parameter regions (1) and (2),
 there exists one de Sitter solution with 
$\tilde{B}_{\rm (dS)}>0$ ($\epsilon=1$). 
In addition, we find three other vacuum solutions 
(two anti de Sitter (AdS) solutions as well as 
a trivial Minkowski spacetime).
We assume $\kappa_f=\kappa_g$.}
\label{epsilon1}
\end{center}
\end{table}

For the coupling constants which satisfy 
\begin{eqnarray*}
&c_3+c_4>0,3c_3+c_4<3,2c_3^2+3c_4>0& [\text{region (3a)}] 
\nn
&c_3+c_4<0,3c_3+c_4>3,2c_3^2+3c_4>0& [\text{region (3b)}] 
\nn
&c_3+c_4>0,3c_3+c_4>3,2c_3^2+3c_4>0& [\text{region (3c)}] 
\,, 
\end{eqnarray*}
we obtain one de Sitter solution for 
$\tilde{B}_{\rm (dS)}<0$ ($B_{\rm (dS)}>0$, $\epsilon=-1$).
In Table \ref{epsilon-1}, we show some examples (Models C, D, and E).
\begin{table}[H]
\begin{center}
  \begin{tabular}{|c|c||c|c|r|r|c|}
\hline 
model&$(c_3, c_4$)&region&$\epsilon$&$B_\ell$~~~~&$\Lambda_g/m_g^2$~~
&vacuum \\ 
\hline 
\hline 
C&   $(1/2, 0)$ &    (3a) & $-1$ & $2+\sqrt{3}$ & $3 \sqrt{3}$ & dS \\
 \cline{4-7}
 &         &       & $-1$ & $2-\sqrt{3}$ & $-3 \sqrt{3} $ &  AdS$_1$   \\
 \cline{4-7}
&         &       &   $1$ & $1$ & $0$ & M \\ \cline{4-7}
 &         &       &   $1$ & $3$ & $-4$ & AdS$_2$    \\ 
\hline \hline
D& $(5/2, -4)$ &    (3b) & $-1$ & $2+\sqrt{3}$ & 
$-72.3731 $ & AdS$_1$  \\ 
\cline{4-7}
 &         &       & $-1$ & $2-\sqrt{3}$ & $0.373067 $ & dS   \\ 
\cline{4-7}
 &         &       &   $1$ & $1/3$ & $-4/9 $ & AdS$_2$  \\ 
\cline{4-7}
 &         &       &   $1$ & $1$ & $0$ & M   \\ \hline \hline
E&   $(3, 0)$ &    (3c) & $-1$ & $0.761557$ & $29.7326$ & dS \\ 
\cline{4-7}
&         &       &   $1$ & $0.636672$ & $-0.154054$ & AdS$_1$     
\\ \cline{4-7}
&         &       &   $1$ & $1$ & $0$ & M \\ \cline{4-7}
&         &       &   $1$ & $4.12489$ & $-23.5786$ & AdS$_2$    
\\ \hline 
 \end{tabular}
    \caption{In the parameter region (3),
 there exists one de Sitter solution with 
$\tilde{B}_{\rm (dS)}<0$ ($\epsilon=-1$). 
We also find three other vacuum solutions 
(two AdS solutions as well as 
a trivial Minkowski spacetime). The region (3) 
is divided into three sub-regions
((3a), (3b) and (3c))
depending on the properties of the solutions.
We assume $\kappa_f=\kappa_g$.}
\label{epsilon-1}
\end{center}
\end{table}

\begin{figure}[h]
\begin{center}
\includegraphics[width=5.5cm,angle=0,clip]{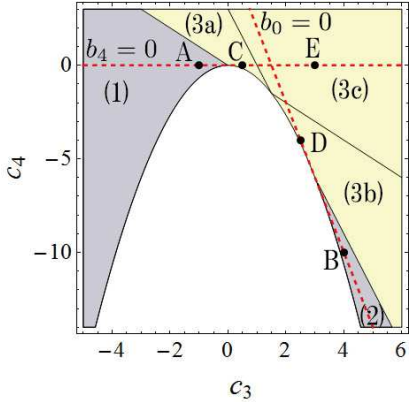}
\caption{The de Sitter 
solutions with $\epsilon=1$ and $\epsilon=-1$ 
are found in the regions (1) and (2) and in the region (3),
respectively. The region (3) is divided into three sub-regions
((3a), (3b) and (3c))
by the dynamical behaviour of the universe.}
\label{deSitter}
\end{center}
\end{figure}

In Fig.\ref{deSitter}, we show the regions (1), (2) and (3) where de Sitter 
solution exists are shown on the $c_3$-$c_4$ plane.
For Models A-E with appropriate coupling parameters, 
we will discuss the dynamics of spacetime later.
They are typical models in each region.

The other two solutions ($\tilde{B}_{\rm (AdS_1 )}$ and $
\tilde{B}_{\rm (AdS_2 )}$), apart from the Minkowski spacetime,
 in the region (1)-(3) are anti de Sitter (AdS)
 spacetimes (see Tables \ref{epsilon1} and \ref{epsilon-1}).

In the white region in Fig.\ref{deSitter},
there is no de Sitter solution. 
There exist either three or one AdS spacetimes.
We show some examples in Table  \ref{sol_B2}.

\begin{table}[H]
\begin{center}
  \begin{tabular}{|c|c||c|c|c|r|c|}
\hline 
Model&$(c_3,c_4)$&region&$\epsilon$&$B$&$\Lambda_g/ m_g^2$~~~&vacuum \\ 
\hline 
\hline 
F&$(-2,-3)$&(4)&$-1$&$0.668907$&$-53.3156$&AdS$_1$  \\ \cline{4-7} 
&&&$1$&$1$&$0$&M \\ \cline{4-7}  
&&&$1$&$1.54181$&$-0.221325$&AdS$_2$  \\ \cline{4-7}  
&&&1&$2.3271$&$-0.183049$&AdS$_3$   \\ 
\hline  \hline
G&$(0,-1)$&(4)&$-1$&$0.537656$&$-15.3417$&AdS$_1$  \\ \cline{4-7} 
&&&$1$&$1$&$0$&M \\ \hline 
 \end{tabular}
    \caption{In the white region of Fig. \ref{deSitter}, 
there is no de Sitter solution.
We find only 
three AdS solutions or one AdS solution 
in addition to a trivial Minkowski spacetime.
We assume $\kappa_f=\kappa_g$.}
\label{sol_B2}
\end{center}
\end{table}

Note that the bare cosmological constant for 
$g_{\mu\nu}$ in the action is $b_0$, but
 the effective 
cosmological constant is given by  $\Lambda_g$ 
through the interaction term. 
Hence, even if $b_0\leq 0$, as long as $\Lambda_g$ is positive, 
we find de Sitter spacetime
as a vacuum solution.

\section{The evolution of the universe}
\label{dynamics}

\subsection{Attractor Universes}
The matter and radiation densities become equal at the redshift 
$z=z_{\rm eq}\approx 3000$ in our universe.
Hence, after $z_{\rm eq}$, matter density is dominant in $g$-spacetime,
which we assume in what follows since we are interested in 
the present acceleration of the universe.
We also assume a flat universe with $k=0$ from observation\footnote{
Even if  we consider the closed ($k=1$) or open ($k=-1$) 
FLRW universe, our main result will not change.}.

We mainly discuss when radiation density in $f$-spacetime 
can be also ignored. In this case, \eqref{A2} becomes
\begin{equation}
C_{\Lambda}(\tilde{B})a_g^3+C_{\text{m}}(\tilde{B})=0
\,,
\end{equation}
which gives 
\bea
a_g(\tilde{B})&=&- \left(\frac{C_{\text{m}}(\tilde{B})}{C_{\Lambda}(\tilde{B})}
 \right)^{\frac{1}{3}} 
\label{ag_B}
\,.
\ena

The potential for $\tilde{B}$ is given by
\begin{eqnarray}
V_g(\tilde{B})=-\frac{3C_{\text{m}}C_{\Lambda}^2
\biggl[ \kappa_g^2{\rho}^{[\gamma]}_g 
C_{\text{m}}
-c_{g,{\rm m}}C_{\Lambda}+3 k
C_{\Lambda}^{\frac{2}{3}}C_{\text{m}}^{\frac{1}{3} }\biggl]}
{(C_{\Lambda}C'_{\text{m}}
-C_{\text{m}}C'_{\Lambda})^2}
\,.
~~~
\label{pot_B}
\end{eqnarray}

If  $C_{\Lambda}(\tilde{B})=0$  as well as $C_{\rm m}(\tilde{B})=0$
 initially,
 $\tilde{B}$ is always constant and then we find the homothetic solution 
as an exact solution: 
\bea
&&
A=B=|\tilde{B}_\ell|
\,,
\nn
&&
c_{f,{\rm m}}=|\tilde{B}_\ell|
c_{g,{\rm m}}
\ena
We find the conventional matter dominant universe with/without 
a cosmological constant\cite{Cosmology4,Anisotropic2}. 

However, $C_{\Lambda}(\tilde{B})$ does not 
usually vanish. 
For generic initial data, 
solving the equation (\ref{eq_B}) for $\tilde{B}$, 
we obtain  the scale factor $a_g$ 
 by Eq. (\ref{ag_B}) with $\tilde{B}(t)$,
 and then another scale factor $a_f$
by
{\setlength\arraycolsep{2pt}\begin{eqnarray}
a_f(\tilde{B})&=& \epsilon \tilde{B}a_g(\tilde{B})
\,. 
\label{b}
\end{eqnarray}}
The ratio $A$ of the lapse functions is also given by 
{\setlength\arraycolsep{2pt}\begin{eqnarray}
A(\tilde{B})&=&\epsilon \left(\tilde{B}
+\frac{3C_{\text{m}}C_{\Lambda}}{C_{\Lambda}C'_{\text{m}}
-C_{\text{m}}C'_{\Lambda}}
\right)\,.
\end{eqnarray}}

\begin{figure}[tbp]
\begin{center}
\includegraphics[width=8cm,angle=0,clip]{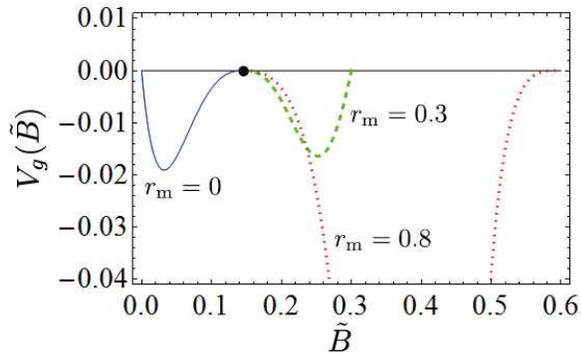}
\caption{The potentials $V_g(\tilde{B})$  for Model B ($c_3=4,c_4=-10$)
with  $r_{\text{m}}=0$ (the blue solid curve), 
$r_{\text{m}}=0.3$ (the green dashed curve), or $r_{\text{m}}=0.8$ 
(the red dotted curve).
 The black dot denotes de Sitter solution $\tilde{B}_{\rm (dS)}$.}
\label{potential}
\end{center}
\end{figure}

The potential $V_g$ satisfies the following conditions
at $\tilde{B}=\tilde{B}_\ell$:
{\setlength\arraycolsep{2pt}\begin{eqnarray}
V_g(\tilde{B}_\ell)&=&0, \\
V'_g(\tilde{B}_\ell)&=&0, \\
V''_g(\tilde{B}_\ell)&=& -6\Lambda_g(\tilde{B}_\ell)
\,.
\end{eqnarray}}
The AdS solution with $\Lambda_g<0$ is isolated 
because the potential is not negative definite
and then Eq.(\ref{eq_B}) is satisfied only at 
$\tilde{B}_\ell=\tilde{B}_{\rm (AdS)}$.
For the case of 
$\Lambda_g>0$, on the other hand, 
there are two allowed regions where the universe can exist;
the left and right regions of the point $\tilde{B}_\ell=
\tilde{B}_{\rm (dS)}$.
The potential near $\tilde{B}_{\rm (dS)}$ 
is shown in Fig. \ref{potential}.
The potential form depends on the ratio of matter densities 
$r_{\rm m}\equiv c_{f,{\rm m}}/c_{g,{\rm m}}$
 as well as the coupling parameters $\{b_i\}$.
Although there are two allowed regions
 in the equation of motion for $\tilde{B}$,
 one side is not physical, that is,
it corresponds to the region where a scale factor becomes negative
because from Eq. (\ref{ag_B}), 
we can evaluate the scale factor near 
$\tilde{B}=\tilde{B}_{\rm (dS)}$ as 
\bea
a_g&=&-\left[{c_{g,{\rm m}}
\tilde{B}_{\rm (dS)}-\epsilon c_{f,{\rm m}}\over 
C'_{\Lambda}(\tilde{B}_{\rm (dS)})
(\tilde{B}-\tilde{B}_{\rm (dS)})}\right]^{1/3}
\nn
&\propto&
 (\tilde{B}-\tilde{B}_{\rm (dS)})^{-1/3}
\ena
which changes the sign at $\tilde{B}=\tilde{B}_{\rm (dS)}$.
Here $C'_{\Lambda}(\tilde{B}_{\rm (dS)})$ is a constant. 
Which side of regions is physical depends on the value of $r_{\rm m}$. 
For example, for Model B ($c_3=4$ and $c_4=-10$), 
if $r_{\rm m}<r_{\rm m}^{\rm (dS)}=0.145979$,
the left region is physically allowed,
 while for $r_{\rm m}>r_{\rm m}^{\rm (dS)}$,
the right region becomes physically possible
(see Fig. \ref{potential}, in which we plot both cases of 
$r_{\rm m}=0$, and $0.3$).

\begin{figure}[h]
\begin{center}
\includegraphics[width=8cm,angle=0,clip]{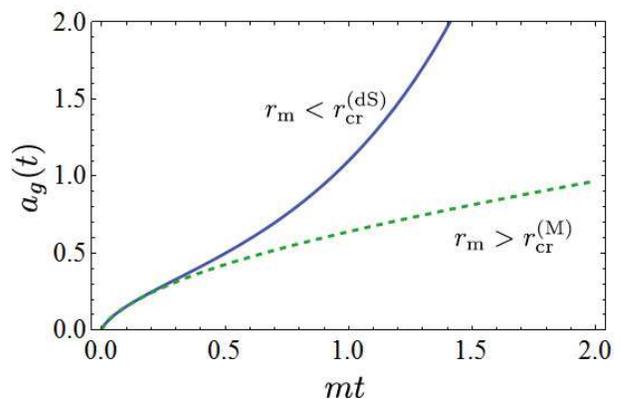}
\caption{The evolution of the scale factor $a_g$ for the case of 
$c_3=4, c_4=-10$.
The bottom curve corresponding to 
$c_{f,{\rm m}}/c_{g,{\rm m}}$
 = 2 (dashed green) 
shows the evolution to dust dominated universe,
while the top curve
corresponding to $c_{f,{\rm m}}/c_{g,{\rm m}}$
 =  0 (solid blue)
shows the evolution to de Sitter spacetime.
}
\label{evolution}
\end{center}
\end{figure}

In both cases, $\tilde{B}$ evolves into $\tilde{B}_{\rm (dS)}$ 
as an attractor.
Near $\tilde{B}_{\rm (dS)}$, the potential is approximated as
\bea
V_{g}\approx -3\Lambda_g (\tilde{B}-\tilde{B}_{\rm (dS)})^2
\,.
\ena
Hence we find the solution for $\tilde{B}$ as
\bea
\tilde{B}\approx \tilde{B}_{\rm (dS)}+C_0 \exp[\pm \sqrt{3\Lambda_g}t]
\,,
\ena
where $C_0$ is an integration constant.
The plus sign corresponds to an unstable evolution rolling down from 
the potential peak, while the minus sign shows 
a stable solution which asymptotically 
approaches to $\tilde{B}_{\rm (dS)}$.
The scale factor evolves as 
\bea
a_g\propto \exp\left[\sqrt{\Lambda_g\over 3}~t\right]
\ena
(see  Fig. \ref{evolution}).
Hence de Sitter accelerating universe is obtained as an attractor.
Note that if $r_{\rm cr}^{\rm (dS)}<r_{\rm m} 
(<r_{\rm m}^{\rm (AdS)}=1.67319)$,
where $r_{\rm cr}^{\rm (dS)}=0.41105$, 
the potential is unbounded from below and diverges at a finite 
value of $\tilde{B}$, where a singularity 
($\dot{\tilde{B}}=\infty$) appears as we will show later
(see the potential with $r_{\rm m}=0.8$ in Fig. \ref{potential}).

For the case of $\tilde{B}_{\rm (M)}=1$, 
we find  
{\setlength\arraycolsep{2pt}\begin{eqnarray}
V_g(\tilde{B}_{\rm (M)})=
V'_g(\tilde{B}_{\rm (M)})=
V''_g(\tilde{B}_{\rm (M)})=0
\,,
\end{eqnarray}}
since $\Lambda_g=0$.
Evaluating also $V'''_g(\tilde{B}_{\rm (M)})$ as
\bea
V'''_g(\tilde{B}_{\rm (M)})&=& 54\left({m_f^2+   m_g^2 r_{\rm m} 
\over 1-  r_{\rm m} }\right)
\,,
\ena
we find that  
$V'''_g$ is positive when $r_{\rm m} <1$,
then the left region ($\tilde{B}\leq \tilde{B}_{\rm (M)}$)
 is physically allowed, 
while the right region ($\tilde{B}\geq \tilde{B}_{\rm (M)}$) 
is allowed  if $r_{\rm m} >1$.

In this case, the potential is approximated as 
\bea
V_g=V_0
(\tilde{B}-\tilde{B}_{\rm (M)})^3
\ena
with
\bea
V_0=9\left({m_f^2+    m_g^2 r_{\rm m}
\over 1-  r_{\rm m} }\right)
\,.
\ena
Eq. (\ref{eq_B}) is integrated as
\bea
-V_0(\tilde{B}-\tilde{B}_{\rm (M)})={4\over (t-t_{0})^2}
\,,
\ena
where 
$t_{0}$ is an integration constant.
As a result, the asymptotic solution of the scale factor is 
\bea
a_g\propto (\tilde{B}-\tilde{B}_{\rm (M)})^{-1/3}\propto
(t-t_{0})^{2/3}
\,,
\ena
which is that of dust matter dominated universe
(see  Fig. \ref{evolution}).
When $\tilde{B}_\ell=\tilde{B}_{\rm (M)}$,
a dust matter dominated universe is found as an attractor.

\subsection{Dynamics of the Universe with Twin Matter}
\label{CNHC}
\begin{figure}[htbp]
\begin{center}
\includegraphics[width=5.5cm,angle=0,clip]{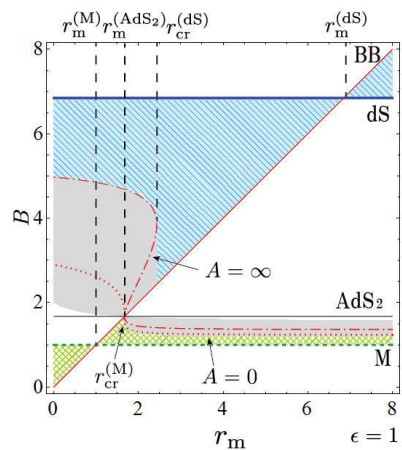}
\caption{The attractor regions in the $r_{\rm m}$-$\tilde{B}$ plane 
are shown for Model A ($c_3=-1,c_4=0$).
The solid, dotted and  dashed  lines denote 
de Sitter, anti de Sitter and dust dominated universes, respectively.
$r_{\rm m}^{\rm (M)}=1$, 
$r_{\rm m}^{\rm (AdS_2)}=1.67319$, and 
$r_{\rm m}^{\rm (dS)}=6.85028$ give the boundary values, 
where the properties of dynamics change.
The initial data in the striped-shaded  light-blue and 
crosshatched light-green regions evolve into 
de Sitter and the dust dominated universe, respectively.
BB denotes an initial Big Bang singularity ($a_g=a_f=0$).
The spacetime started from the other colored region
evolves into singularities, which are shown by dot-dashed curves. 
There exist two critical values $r_{\rm cr}^{\rm (dS)}=2.4328$ 
and $r_{\rm cr}^{\rm (M)}=1.67318$. 
Beyond $r_{\rm cr}^{\rm (dS)}$, 
every spacetime evolves into de Sitter universe 
if $B>r_{\rm m}^{\rm (AdS_2)}$, 
while all spacetime with 
$r_{\rm m}<r_{\rm cr}^{\rm (M)}$ 
evolves into the matter dominant universe 
if $B<r_{\rm m}^{\rm (AdS_2)}$. }
\label{region1}
\end{center}
\end{figure}

\begin{figure}[htbp]
\begin{center}
\includegraphics[width=5.5cm,angle=0,clip]{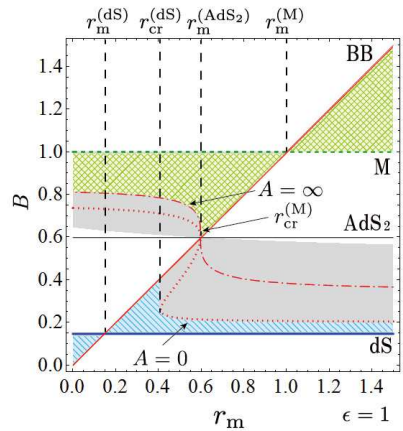}
\caption{The same figure as Fig. \ref{region1}
for  Model B ($c_3=4,c_4=-10$). The boundary values are given by 
$r_{\rm m}^{\rm (M)}=1$, 
$r_{\rm m}^{\rm (AdS_2)}=0.59766$, and 
$r_{\rm m}^{\rm (dS)}=0.145979$. 
Below the critical value $r_{\rm cr}^{\rm (dS)}=0.41105$,
every spacetime evolves into de Sitter universe 
if $B<r_{\rm m}^{\rm (AdS_2)}$, while all
 spacetime with $r_{\rm m}>r_{\rm cr}^{\rm (M)}=0.597663$ 
evolves into the matter dominant universe if 
$B>r_{\rm m}^{\rm (AdS_2)}$.
} 
\label{region2}
\end{center}
\end{figure}

\begin{figure}[htbp]
\begin{center}
\includegraphics[width=6cm,angle=0,clip]{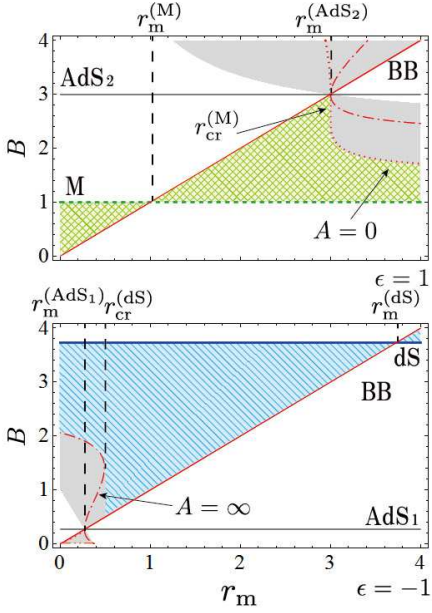}
\caption{The same figure as Fig. \ref{region1}
for  Model C ($c_3=1/2,c_4=0$). 
The de Sitter solution exists in the case of $\epsilon=-1$, 
while the matter dominant universe is found for $\epsilon=1$.
Hence we draw two figures of $\epsilon=\pm 1$ separately. 
because there appears a singularity at $\tilde B=0$,
where $\epsilon$ changes the sign.
The boundary values are given by 
$r_{\rm m}^{\rm (M)}=1$, 
$r_{\rm m}^{\rm (AdS_1)}=2-\sqrt{3}$,
$r_{\rm m}^{\rm (AdS_2)}=3$, and 
$r_{\rm m}^{\rm (dS)}=2+\sqrt{3}$. 
Beyond the critical value $r_{\rm cr}^{\rm (dS)}=0.489757$,
every spacetime evolves into de Sitter universe if 
$\epsilon=-1$ and $B>r_{\rm m}^{\rm (AdS_1)}$,
 while all spacetime  with $r_{\rm m}<r_{\rm cr}^{\rm (M)}=2.99645$ 
evolves into the matter dominant universe 
if $\epsilon=1$ and $B<r_{\rm m}^{\rm (AdS_2)}$.}
\label{region3a}
\end{center}
\end{figure}

\begin{figure}[htbp]
\begin{center}
\includegraphics[width=6cm,angle=0,clip]{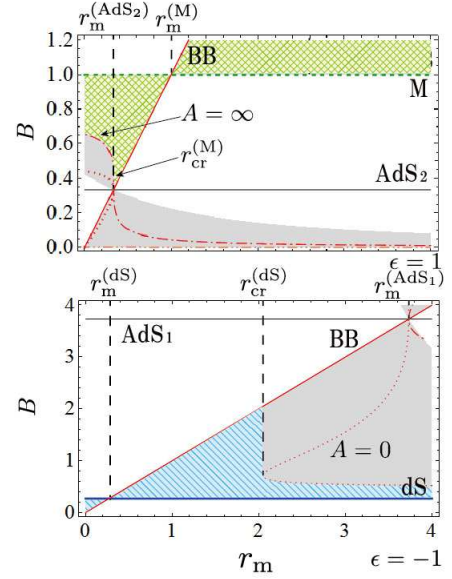}
\caption{The same figure as Fig. \ref{region3a}
for  Model D ($c_3=5/2,c_4=-4$) ($\epsilon=-1$). 
The boundary values are given by 
$r_{\rm m}^{\rm (M)}=1$, 
$r_{\rm m}^{\rm (AdS_1)}=2+\sqrt{3}$,
$r_{\rm m}^{\rm (AdS_2)}=1/3$, and 
$r_{\rm m}^{\rm (dS)}=2-\sqrt{3}$. 
Below the critical value $r_{\rm cr}^{\rm (dS)}=2.04183$,
every spacetime evolves into de Sitter universe 
if $\epsilon=-1$ and $B<r_{\rm m}^{\rm (AdS_1)}$, 
while all spacetime  with 
$r_{\rm m}>r_{\rm cr}^{\rm (M)}=0.33729$ evolves into 
the matter dominant universe
if $\epsilon=1$ and $B>r_{\rm m}^{\rm (AdS_2)}$.}
\label{region3b}
\end{center}
\end{figure}

\begin{figure}[htbp]
\begin{center}
\includegraphics[width=6cm,angle=0,clip]{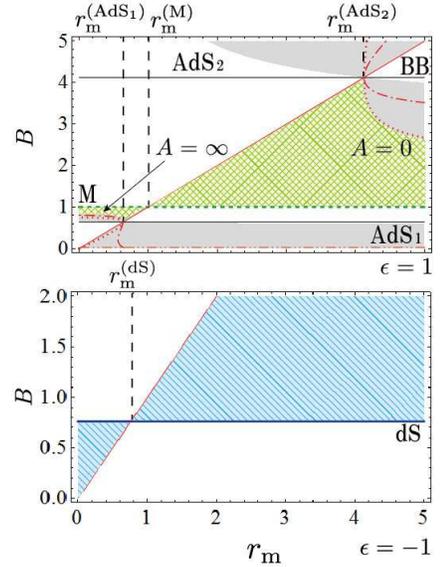}
\caption{The same figure as Fig. \ref{region3a}
for  Model E ($c_3=3,c_4=0$). The boundary values are given by 
$r_{\rm m}^{\rm (M)}=1$, 
$r_{\rm m}^{\rm (AdS_1)}=0.636672$,
$r_{\rm m}^{\rm (AdS_2)}=4.12489$, and 
$r_{\rm m}^{\rm (dS)}=0.761557$. 
Every spacetime with $\epsilon=-1$ 
evolves into de Sitter universe.
The matter dominant universe is found for 
 all spacetime 
with 
$\epsilon=1$ and $r_{\rm m}^{\rm (AdS_1)}<
r_{\rm m}<r_{\rm m}^{\rm (AdS_2)}$
}
\label{region3c}
\end{center}
\end{figure}

We are interested in whether the cosmic no-hair conjecture 
holds. Hence we analyze our system for various initial data
and discuss which initial condition leads to de Sitter expansion.
In order to discuss whether de Sitter accelerating universe 
is naturally achieved as an attractor or not, we survey all 
 possible allowed initial data.
Especially we focus on the ratio $r_{\rm m}$ 
of energy densities of twin matter fluids. 
The results are summarized on the $r_{\rm m}$-$\tilde{B}$  plane. 
For the parameter region (1) and (2),
we show two typical examples of Model A($c_3=-1, c_4=0$)
and of Model B($c_3=4, c_4=-10$), 
in Figs. \ref{region1} and \ref{region2}, respectively.
For the region (3), 
we also present the typical results for Model C($c_3=1/2, c_4=0$),
D ($c_3=5/2, c_4=-4$)
and E ($c_3=3, c_4=0$) in Figs. \ref{region3a},
 \ref{region3b} and \ref{region3c},
respectively.

The colored regions denote the ranges of physically allowed initial data. 
The universes in the stripe-shaded light-blue area evolve into de Sitter 
spacetime, while those in the  crosshatched light-green area evolve into 
the dust matter dominated universe.
The universes started from the 
grey shaded areas eventually find a future singularity.


\begin{figure}[h]
\begin{center}
\includegraphics[width=6.5cm,angle=0,clip]{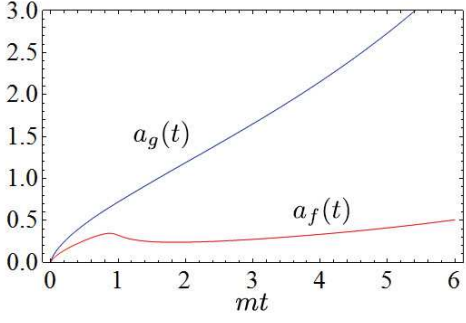}
\caption{The time evolution of two scale factors
$a_g$ and $a_f$ for Model B ($c_3=4,c_4=-10$) with
$c_{f,{\rm m}}/c_{g,{\rm m}}=0.58$.}
\label{agt}
\end{center}
\end{figure}

\begin{figure}[h]
\begin{center}
\includegraphics[width=6.5cm,angle=0,clip]{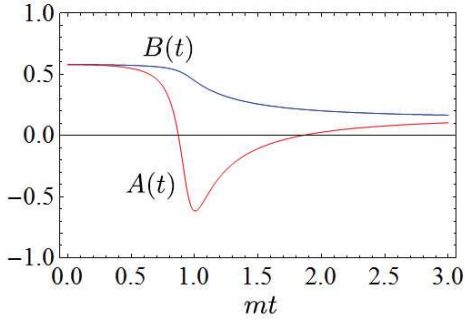}
\caption{The time evolution of $B(t)$ and $A(t)$ 
for Model B   with
$c_{f,{\rm m}}/c_{g,{\rm m}}=0.58$.}
\label{AtBt}
\end{center}
\end{figure}

\begin{figure}[h]
\begin{center}
\includegraphics[width=6.5cm,angle=0,clip]{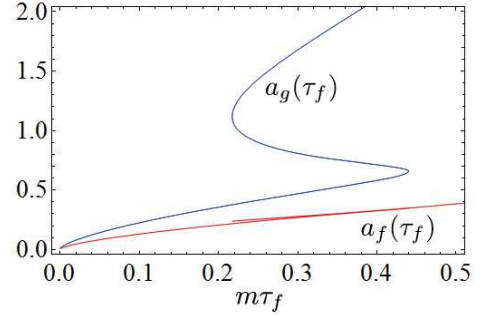}
\caption{The time evolution of two scale factors
$a_g$ and $a_f$ in terms of $\tau_f$
 for Model B   with
$c_{f,{\rm m}}/c_{g,{\rm m}}=0.58$.
Beyond the singularity, both scale factors decrease in time,
and then increase again after another singularity.}
\label{agtau}
\end{center}
\end{figure}

We may probably easily understand that the spacetime evolves 
either de Sitter universe or the matter dominant universe,
depending on the initial conditions, 
because two homothetic solutions are attractors.
However we also find singular spacetime for some initial data.
Why a flat FLRW universe can evolves into 
a future singularity, which never happens in GR ?
To explain how the universe evolves into a singularity,
 we consider Model B ($c_3=4$ and $c_4=-10$)
(Fig. \ref{region2}).
If $r_{\rm m}<r_{\rm cr}^{\rm (dS)}$, 
the universe starts from a big bang initial singularity 
and evolves into de Sitter spacetime.
When $r_{\rm cr}^{\rm (dS)}<r_{\rm m}<r_{\rm m}^{\rm (AdS_2)}$, 
the evolution of $a_g$ is the similar to the above case.
Starting from a big bang initial data ($a_g=0$), it evolves into de Sitter 
spacetime. However the behaviour of $a_f$ becomes strange.
We show the time evolution of two scale factors in Figs. 
\ref{agt}-\ref{agtau}, in which 
we set $r_{\rm m}=0.58$. 

$a_f$ first increases and then turns to decrease.
It eventually increases again, resulting in 
an exponential expansion. 
In order to analyze the reason why the universe shows 
a transient collapse, we show the time evolution of 
$A$ and $B$ in Fig. \ref{AtBt}.
We find that $A$ becomes negative when $a_f$ decreases.
It means that the time direction in this period 
turns to be reverse. It is the reason of the collapse.

However there appears a singularity when $\tilde{A}$ vanishes,
 i.e., $\dot a_f=0$.
Substituting 
$\tilde{A}=\dot{a}_f/\dot{a}_g$ into the Ricci scalar 
$\mathcal{R}(f)$, we obtain
\begin{eqnarray}
\mathcal{R}(f)&=&
6\left[{1\over N_f a_f}\left({\dot a_f \over N_f}\right)^{\cdot}
+\frac{\dot{a}_f^2}{N_f^2 a_f^2}+\frac{k}{a_f^2} \right]
\nn
&=&
6\left(\frac{\dot{a}_g \ddot{a}_g}{a_f \dot{a}_f}
+\frac{\dot{a}_g^2}{a_f^2}+\frac{k}{a_f^2} \right)\,.
\end{eqnarray}
$\ddot{a_g}$ does not vanish at $\tilde{A}=0$,
 because the $r$-$r$ component 
of field equation of $g_{\mu\nu}$ is given by
{\setlength\arraycolsep{2pt}\begin{eqnarray}
&&2\frac{\ddot{a_g}}{a_g} + \frac{\dot{a_g}^2}{a_g^2} +\frac{k}{a_g^2}  
={m_g^2}\Big[b_0+b_1(\tilde{A}+2\tilde{B})
\nn
&&
~~~~
+b_2(2\tilde{A}\tilde{B}+\tilde{B}^2)+b_3\tilde{A}\tilde{B}^2\Big]
- \kappa _g^2 P_g 
\,.\label{g2} 
\end{eqnarray}}
Thus, the Ricci scalar $\mathcal{R}(f)$ diverges at $\dot{a}_f=0$ 
($\tilde{A}=0$) assuming $\dot{a}_g \neq 0$. 
Note that the Ricci scalar for $g_{\mu\nu}$ is finite 
at this point. 
It implies that $g$-spacetime  is regular whereas $f$-spacetime  
is singular at $\tilde{A}=0$.

Note that 
it is possible to solve the equation for $\tilde{B}$  by use of 
the potential $V_g$ even we find a singularity in $f$-spacetime,
because the interaction term does not diverge. 
However, it is impossible to solve the equation 
by use of the cosmic time $\tau_f$ in $f$-spacetime.
From Fig. \ref{agtau}, it is almost trivial that 
a singularity appears at the turning points of $a_f$.

Fig.\ref{AtBt} implies that  
$t(\tau_f)$ is not single-valued although 
$\tau_f(t)$ is a single-valued function. As a result,
the variables such as $a_g(\tau_f)$ or $a_f(\tau_f)$
 are not single-valued
(see Fig.\ref{agtau}). 

\begin{figure}[h]
\begin{center}
\includegraphics[width=6.5cm,angle=0,clip]{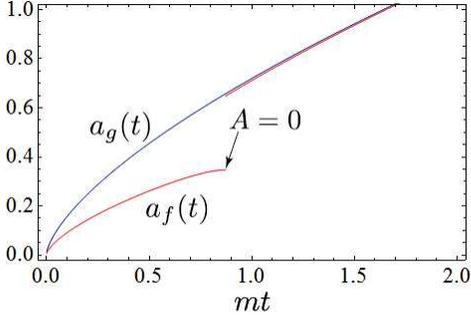}
\caption{The time evolution of two scale factors
 for the same parameters as those in Fig. \ref{agt}.
Although both scale factors increase in time,  
there still exists a singularity at $A=0~(\dot a_f=0)$
and $a_f$ is discontinuous there.
The parity of  $f$-spacetime 
is changed beyond the singularity. 
}
\label{agt2}
\end{center}
\end{figure}

Beyond a singularity, however, 
since there is no natural junction condition at 
the singularity, 
we can change the sign of the lapse function
$N_f$, which is negative when $A$ is negative.
Since the tetrad in $f$-spacetime is given by 
(\ref{tetrad_f}), if we change the sign of $N_f$, 
we also have to change the sign of the spatial part.
Because the scale factor $a_f$ must be positive, 
we have to reverse the spatial direction, that is, 
we should change the parity in $f$-spacetime.
Hence if we keep the time direction in $f$-spacetime beyond 
a singularity, 
we have to change the parity for the period of $A<0$.
Although there is no contraction phase in $f$-spacetime as well as 
in $g$-spacetime, the scale factor in $f$-metric $a_f$ becomes 
discontinuous (see Fig. \ref{agt2}).

The reverse of the above case also occurs, that is,
  $f$-spacetime is regular anytime except for a big bang singularity
 whereas $g$-spacetime becomes singular at $\dot{a}_g=0$,
when  $\tilde{A}=\infty$.
The sign of $\tilde{A}$ also changes beyond this singularity.
This happens in the case of Model A ($c_3=-1, c_4=0$).

In Figs. \ref{region1}-\ref{region3c}, 
we show the region of $\tilde{A}< 0$, on which boundaries 
(the solid and dashed curves for  $\tilde{A}=0$ $\tilde{A}=\infty$,
 respectively) singularities appear.
Hence if the universe starts from the gray shaded area,
it evolves into a singularity either at $\tilde{A}=0$ or 
at  $\tilde{A}=\infty$.
 If the universe starts from a big bang singularity ($a_g=a_f=0$: 
red solid line), 
it evolves into a negative lapse area through a singularity 
and eventually goes to a positive lapse area again, finding 
de Sitter accelerating universe
(or the matter dominated universe).
For the other initial data in  the grey area,
 the boundary does not correspond to a big bang singularity,
but the universe is bounced at the boundary.
Either this spacetime evolves directly into a singularity 
at $\tilde{A}=\infty$, 
or it first goes to the boundary and then it is 
bounced back to the singularity.
Going through a negative lapse area, both cases
 eventually evolve into a positive lapse area again.
In any case, however, a singularity formation 
cannot be avoided if the universe starts from the grey area.

As shown in Figs. \ref{region1}-\ref{region3c}, 
there exists critical values $r_{\rm cr}^{\rm (dS)} ( r_{\rm cr}^{\rm (M)})$  for $r_{\rm m}=
c_{f,{\rm m}}/c_{g,{\rm m}}$, beyond (or below) which 
both $g$- and $f$-spacetime are regular and then they evolve into 
de Sitter universe (or the dust matter dominated universe).
The universe never evolves into a singularity.

The critical value 
 $r_{\rm cr}^{\rm (dS/M)}$ can be found as follows:
It is given by an extreme value of 
boundary curve of $\tilde{A}=0$ or $\tilde{A}=\infty$,
which are given by 
\begin{eqnarray}
r_{\rm m}|_{A=0}&=&{c_{f,{\rm m}}\over c_{g,{\rm m}}}\Big{|}_{A=0}
=
\epsilon 
\tilde{B}\left(1+{C_\Lambda\over 3C_\Lambda-\tilde B C'_{\Lambda}}\right)
\,,~~~~ 
\label{A=0}
 \\
r_{\rm m}|_{A=\infty}&=&{c_{f,{\rm m}}
\over c_{g,{\rm m}}}\Big{|}_{A=\infty}
=
\epsilon \left(\tilde{B}-{C_\Lambda\over  C'_\Lambda}\right)
\,.
 \label{A=infty}
\end{eqnarray}
The extremal condition gives the equation for 
$\tilde B$ at the critical point such that 
\begin{eqnarray*}
(\kappa_g^2 b_1-\kappa_f^2b_3)\tilde{B}^2
+(\kappa_g^2 b_0-3\kappa_f^2b_2)\tilde{B}
-2\kappa_f^2b_1&=&0
\,,
\end{eqnarray*}
for $ \tilde A=0$ or 
\begin{eqnarray*}
2\kappa_g^2b_3\tilde{B}^2+(3\kappa_g^2b_2-\kappa_f^2b_4)\tilde{B}
+(\kappa_g^2b_1-\kappa_f^2b_3)
&=&0
\end{eqnarray*}
for $ \tilde A=\infty$, respectively. 
The roots of the above equation just provide 
a candidate for the critical value $\tilde{B}_{{\rm cr}}$.
Since the critical point must exist in the physically allowed region, 
we have to impose  
the additional constraint for the critical value as 
\bea
V_g(\tilde{B}_{{\rm cr}},r_{\rm cr}^{\rm (dS/M)})<0 \label{additional_constraint}
\,.
\ena
for  $\tilde{B}_{\rm (dS/M)}
>\tilde{B}_{{\rm cr}}>r_{\rm m}$ 
or 
$\tilde{B}_{\rm (dS/M)}
<\tilde{B}_{{\rm cr}}<r_{\rm m}$.
These critical values $r_{\rm cr}^{\rm (dS/M)}$ 
are shown in Figs. \ref{region1}-\ref{region3c}. 

We summarize the results in this subsection in Table \ref{summary2}.
\begin{table}[h]
\begin{center}
\begin{tabular}{|c||c|c|c|}
\hline 
region&$\epsilon$&$r_{\rm m}$&$B$
\\ 
\hline 
\hline 
\multicolumn{4}{|c|}{de Sitter accelerating universe} 
\\ 
\hline
(1) &$1$&$r_{\rm m}>r_{\rm cr}^{\rm (dS)}$&$B>r_{\rm m}^{\rm (AdS_2)}$
\\
\hline 
(2) &$1$&$r_{\rm m}<r_{\rm cr}^{\rm (dS)}$&$B<r_{\rm m}^{\rm (AdS_2)}$
\\
\hline 
(3a) &$-1$&$r_{\rm m}>r_{\rm cr}^{\rm (dS)}$&$B>r_{\rm m}^{\rm (AdS_1)}$
\\
\hline 
(3b) &$-1$&$r_{\rm m}<r_{\rm cr}^{\rm (dS)}$&$B<r_{\rm m}^{\rm (AdS_1)}$
\\
\hline 
(3c) &$-1$&no condition &no condition
\\ 
\hline 
\hline 
 \multicolumn{4}{|c|}{matter dominant universe} 
\\
\hline 
(1) &$~1~$&$r_{\rm m}<r_{\rm cr}^{\rm (M)}$&$B<r_{\rm m}^{\rm (AdS_2)}$
\\
\hline 
(2) &$1$&$r_{\rm m}>r_{\rm cr}^{\rm (M)}$&$B>r_{\rm m}^{\rm (AdS_2)}$
\\
\hline 
(3a) &$1$&$r_{\rm m}<r_{\rm cr}^{\rm (M)}$&$B<r_{\rm m}^{\rm (AdS_2)}$
\\
\hline 
(3b) &$1$&$r_{\rm m}>r_{\rm cr}^{\rm (M)}$&$B>r_{\rm m}^{\rm (AdS_2)}$
\\
\hline 
(3c) &$1$&$r_{\rm cr}^{\rm (AdS_1)}<r_{\rm m}<r_{\rm cr}^{\rm (AdS_2)}$&
$r_{\rm cr}^{\rm (AdS_1)}<B<r_{\rm cr}^{\rm (AdS_2)}$
\\ \hline 
\end{tabular}
\caption{The conditions for de Sitter accelerating universe
or the matter dominant universe.
Every spacetime  evolves into de Sitter universe 
or the matter dominant universe, if the given 
conditions are satisfied 
for $\epsilon, r_{\rm m}$ and the initial value of $B$. }
\label{summary2}
\end{center}
\end{table}

We show the conditions for the ratio $r_{\rm m}$ and the initial value
of $B$ under which  every spacetime  evolves into de Sitter universe
or the matter dominant universe.
The critical values depend on the coupling constants $\{b_i \}$
(or $c_3$ and $c_4$)
 and $\kappa_f^2/\kappa_g^2$.
When spacetimes do not satisfy these conditions, 
the universe will find a singularity unless 
we fine-tune the initial conditions.

\subsection{Cosmic No Hair Conjecture}
\label{cosmic_no_hair}
In the previous subsection, 
we discuss several examples, in which 
we showed that there are three possibilities for the fate of 
spacetime: de Sitter accelerating universe, the matter dominant 
universe, and spacetime with a future singularity, depending on 
the initial condition.
Hence in the exact sense, the cosmic no hair conjecture 
does not hold, but de Sitter universe can be obtained 
from a wide range of initial conditions. 
In this subsection, we shall further 
analyze how this result is generic by surveying the possible
coupling parameters $\{b_k\}$, which are 
given by two free parameters $c_3$ and $c_4$ as 
\eqref{flatsol}. 

\begin{figure}[h]
\begin{center}
\includegraphics[width=6cm,angle=0,clip]{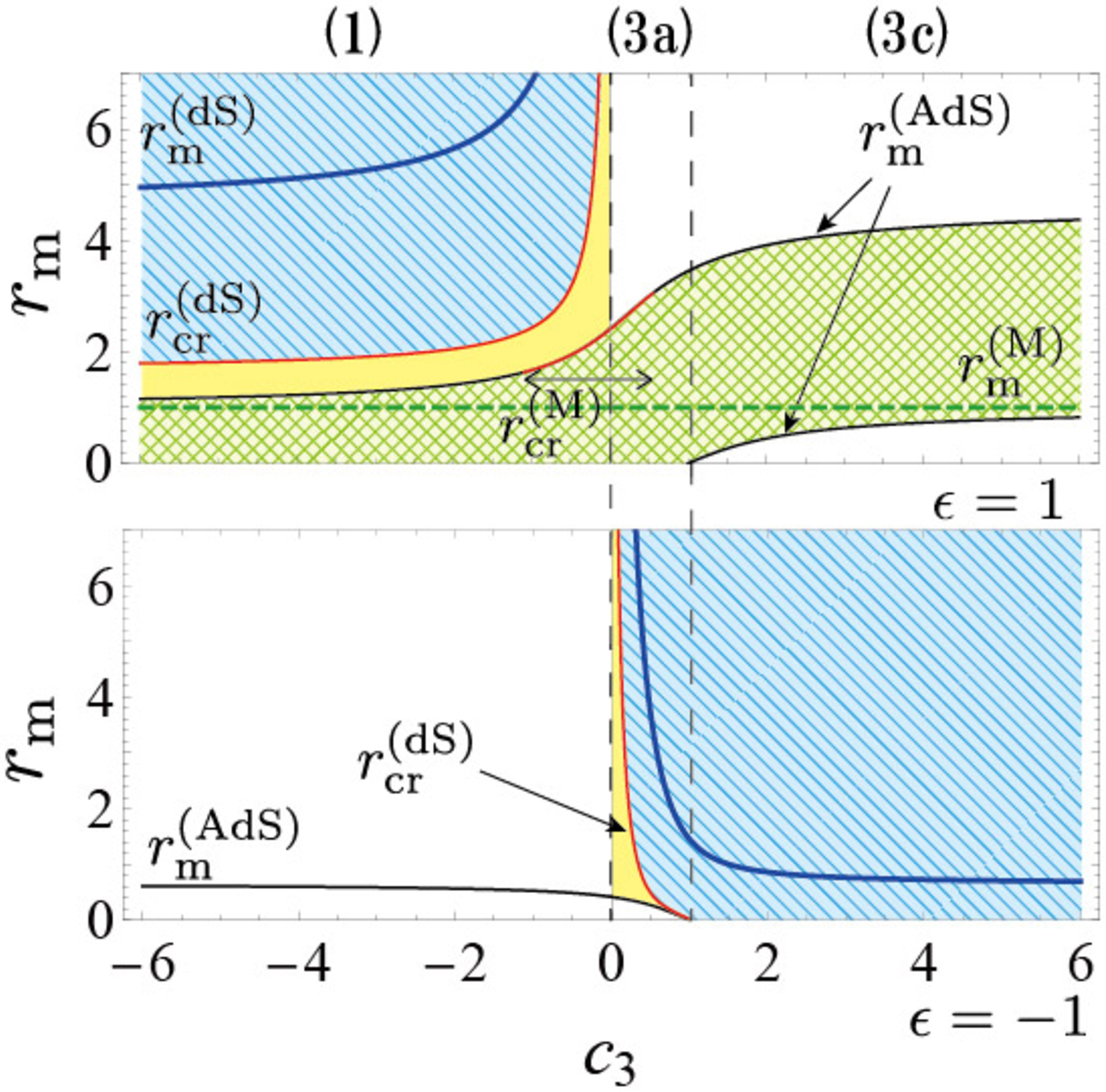}
\\
(a) Case (I)
\\
\includegraphics[width=6cm,angle=0,clip]{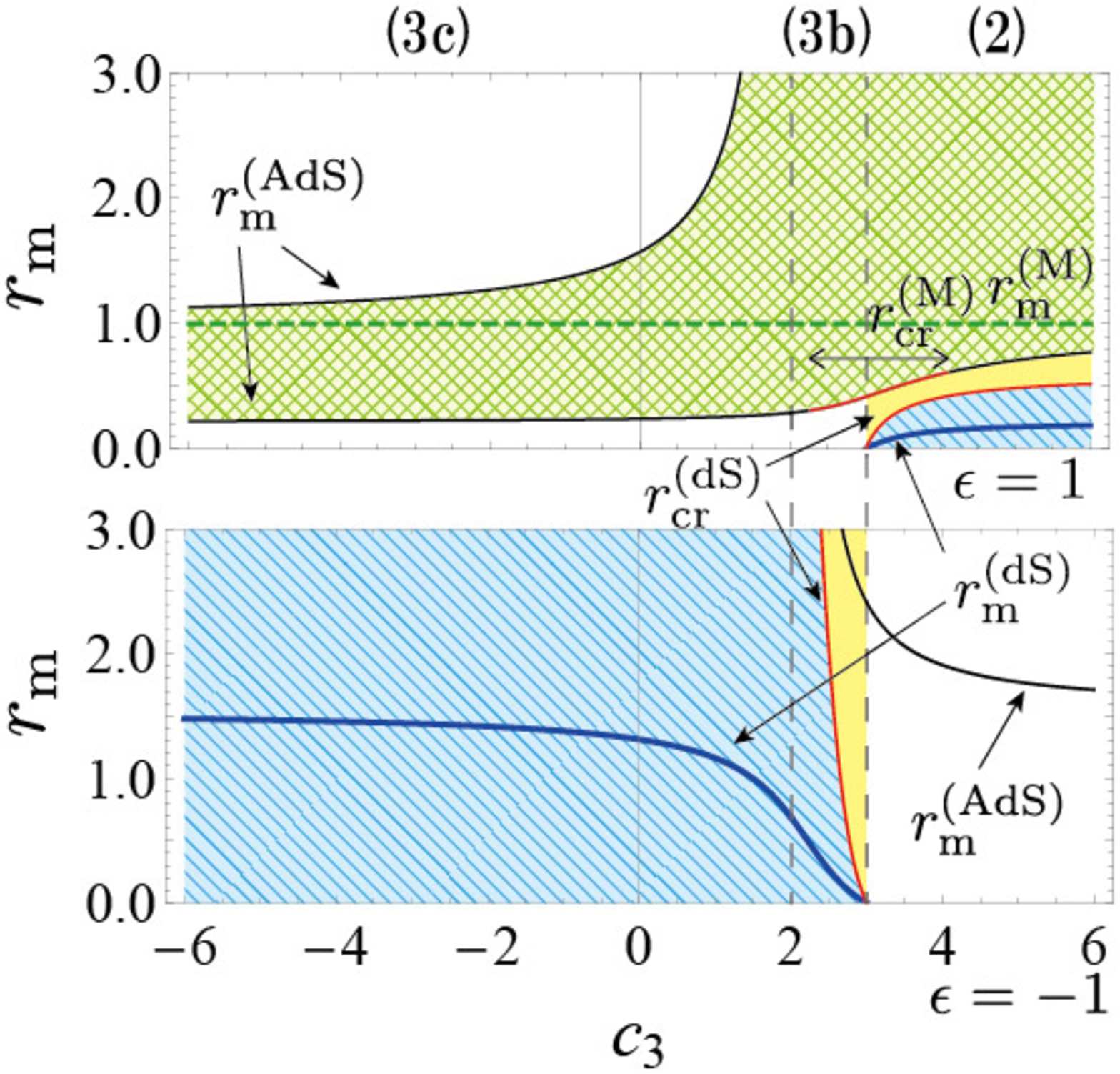}
\\
(b) Case (II)
\\
\caption{The de Sitter solution  (blue curves) and the necessary condition 
 of $r_{\text{m}}$ for self-acceleration (the stripe-shaded light-blue regions) are shown. 
We also shown the matter dominate universe 
(the crosshatched light-green curve) and  
its necessary condition 
 of $r_{\text{m}}$.
The dashed red 
curves show AdS solutions with $\Lambda_g<0$. 
If the universe starts from the yellow region,
it evolves into a singularity. 
The critical value $r_{\rm cr}^{\rm (dS)}$ 
exists in the regions (1), (2), (3a) and (3b).
The another critical value
 $r_{\rm cr}^{\rm (M)}$ appear if $-1.09<c_3<0.55$ for Case (I) 
and $2.27<c_3<4.09$ for Case (II), respectively.}
\label{b0=0}
\end{center}
\end{figure}

Here, just for simplicity, we study two 
typical cases with one free parameter $c_3$:
(I)  $b_4=c_4=0$ and (II) $b_0=4c_3+c_4-6=0$. 
The first and second cases include  the region (1), (3a) and (3c), 
and  the region (2), (3b) and (3c), respectively.
(See the corresponding red dashed lines in Fig.\ref{deSitter}.)

In Figs. \ref{b0=0} (a) and (b),  for 
those two cases (I)  and (II), 
we show which range of 
$r_{\rm m}$ can reach to de Sitter universe 
or the matter dominant universe, 
The blue solid curve and green dashed line denote de Sitter solution 
and the matter dominated universe, respectively.
For the dS solutions, 
the value of $\epsilon$ is negative in the regions (3a), (3b) and (3c)
while it is positive in the regions (1) and (2).

In the region (3c), all spacetime evolves into either de Sitter 
self-accelerating universe 
or matter dominant universe, except for the time reversed ones,
which collapse into a big crunch.
On the other hand, 
in the regions (1), (2), (3a) and (3b),
there exists a critical value $r_{\rm cr}^{\rm (dS)}$,
beyond (below) which spacetime with an appropriate initial condition 
evolves into de Sitter universe.
However the case with other initial data will evolve into 
either matter dominant universe or find a singularity.
We note the critical values $r_{\rm cr}^{\rm (M)}$ are 
extremely close to $r_{\rm m}^{\rm (AdS)}$ 
and these appear only if $-1.09<c_3<0.55$ for Case (I) and 
$2.27<c_3<4.09$ for Case (II).
Outside these regions, magnitude relation between $B_{\rm cr}$ 
and $r_{\rm m}$ is $B_{\rm cr}>r_{\rm m}>r_{\rm m}^{\rm (M)}$ 
or $B_{\rm cr}<r_{\rm m}<r_{\rm m}^{\rm (M)}$, which dose not 
satisfy the additional constraint \eqref{additional_constraint}.

\begin{figure}[h]
\begin{center}
\includegraphics[width=6cm,angle=0,clip]{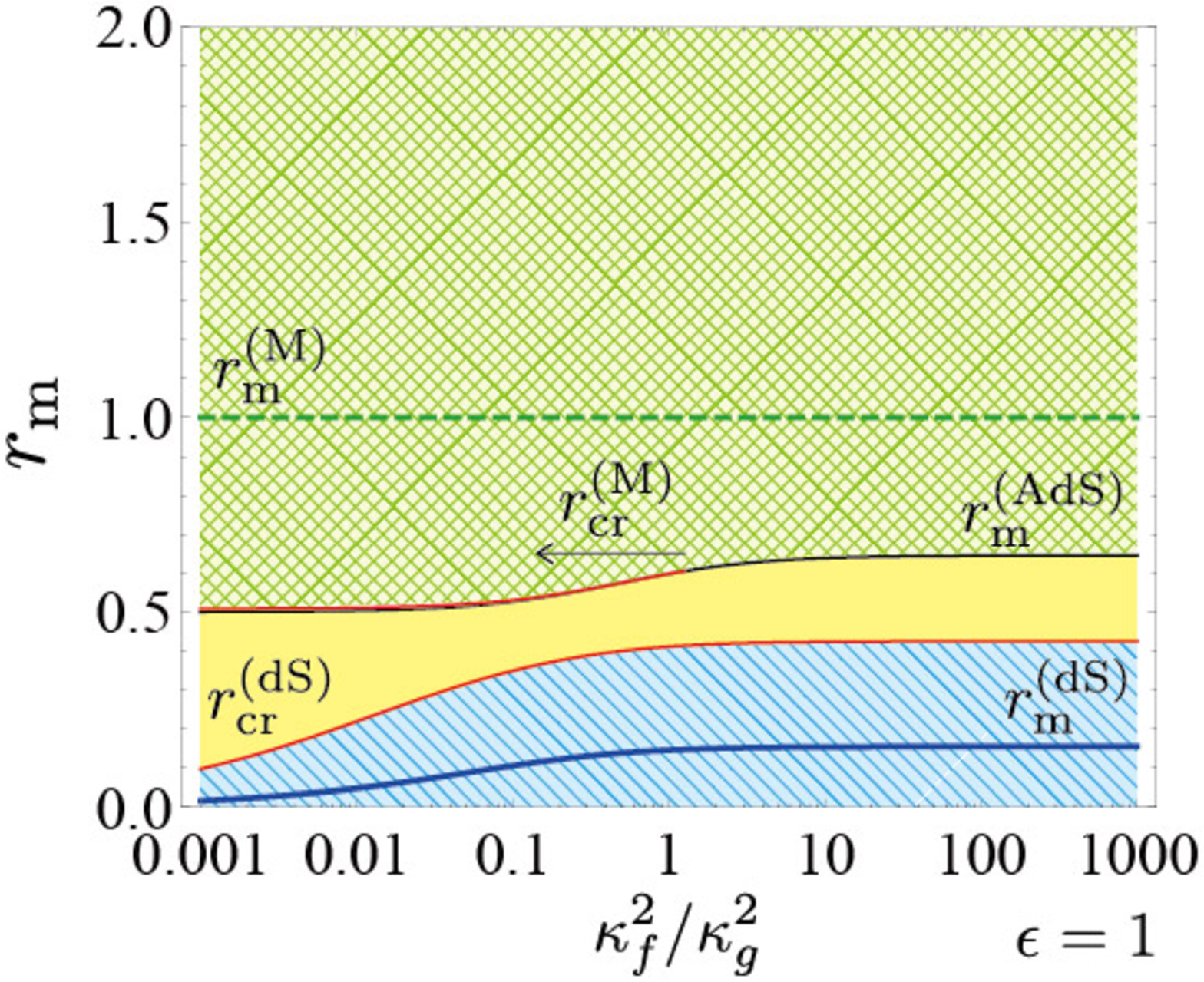}
\\
(a)
\\
\includegraphics[width=6cm,angle=0,clip]{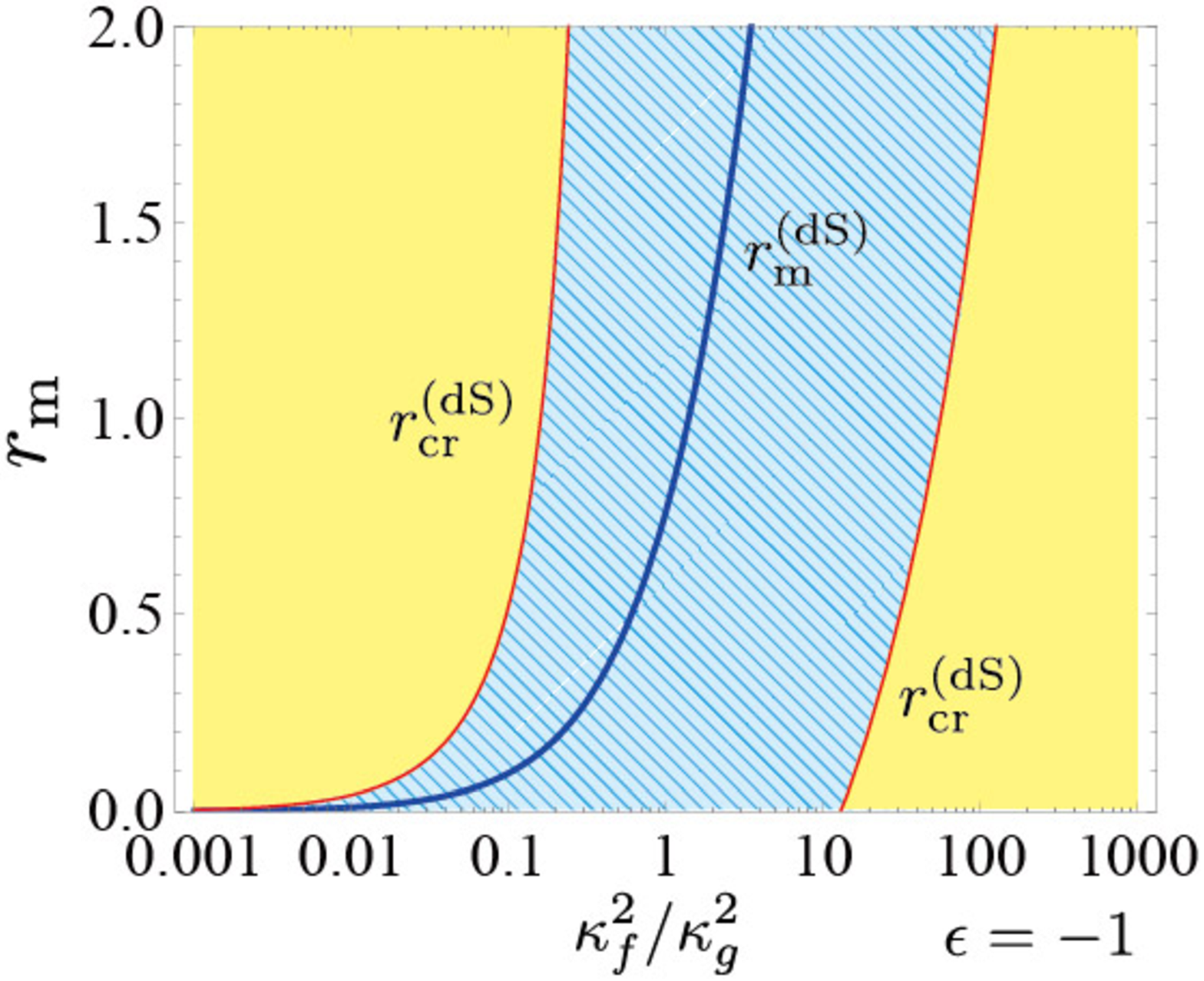}
\\
(b)
\\
\caption{The similar figures to Fig. \ref{b0=0} 
for different values of $\kappa^2_f/\kappa^2_g$.
We plot the results for 
(a) Model B ($c_3=4,c_4=-10$) and  (b) Model E ($c_3=3,c_4=0$).
For Model B, in addition to the critical value  $r_{\rm cr}^{\rm (dS)}$
another critical value $r_{\rm cr}^{\rm (M)}$ appears 
if $\kappa_f^2/\kappa_g^2<1.26$.
For Model E, no critical value appears 
if  $0.522408<\kappa_f^2/\kappa_g^2<13.0711$.
 All spacetime approach de Sitter universe.}
\label{kappa}
\end{center}
\end{figure}

In order to see the dependence of gravitational constants 
$\kappa_g$ and $\kappa_f$, 
we change the ratio of gravitational constants $\kappa_f/\kappa_g$
by fixing the coupling constants $\{b_i\}$.
In Figs.\ref{kappa} (a) and (b), we show 
the results for different values of the ratio 
$\kappa^2_g/\kappa^2_g$ 
for Model B ($c_3=4,c_4=-10$) and Model E ($c_3=3,c_4=0$), 
respectively.

The result is qualitatively same in Model B except for the existence of 
$r_{\rm cr}^{\rm (M)}$, while it is quite different in Model E. 
For Model B, the critical value $r_{\rm cr}^{\rm (M)}$ 
appears only if $\kappa_f^2/\kappa_g^2<1.26$ as shown in Fig. \ref{b0=0}. 
For Model E, all spacetime evolves into de Sitter universe
 if $0.522408<\kappa_f^2/\kappa_g^2<13.0711$,
otherwise a critical value appears as Model B.

Hence we can conclude that 
no hair conjecture does not always hold in the exact sense, 
but a self-accelerating universe can be found from 
natural (not fine-tuned) initial data for general coupling parameters 
and gravitational constants.

We should note the effect of radiation.
Although the present radiation density is much less than 
matter density in our universe, 
it may not  be the case for $f$-matter fields.
Analyzing the case that 
$f$-radiation density is not ignorable, we 
find that 
the dynamics of the universe does not 
change so much from the matter dominated case,
although the interpretation of ``dark matter'' component 
will be different (see \S. \ref{dark_radiation}).

\section{Toward the $\Lambda$-CDM universe}\label{observation}
\subsection{Effective Friedmann Equation}
As we show, de Sitter accelerating universe is realized 
for generic initial conditions in bi-gravity theory.
One may wonder whether our present observed universe is 
found in this model.
Among many cosmological models, the $\Lambda$-CDM model is 
most preferable from the observational view point \cite{Planck}.
The amount of cold dark matter is five times as large as 
the baryonic matter. 
Can we obtain such a model in the present theory 
as an attractor or without any fine-tuning ?

In order to study it, assuming the spacetime approaches 
the homothetic solution ($\tilde B=\tilde B_\ell$), 
we describe our basic equation (\ref{g1})
in the form of a standard Friedmann equation.
Since de Sitter spacetime or the matter dominant universe 
is an attractor in the present model, 
$\tilde B_\ell=\tilde B_{\rm (dS)}$ or $\tilde B_{\rm (M)}$.

We rewrite the interaction term $\rho^{[\gamma]}_g$ 
in terms of the energy densities of twin matter fluids, $\rho_g$ 
and $\rho_f$.
Near $\tilde B=\tilde B_\ell$, 
this term is expanded as \\[.5em]
\bea
\kappa_g^2 \rho^{[\gamma]}_g\approx \Lambda_g+
R_1 (\tilde B-\tilde B_\ell)+O((\tilde B-\tilde B_\ell)^2)
\label{rho_g_gam_expand}
\,,
\ena
where 
\bea
R_1\equiv 
\Big(\kappa_g^2 \rho^{[\gamma]}_g\Big)'(\tilde B_\ell)
=3m_g^2\left(b_1+2b_2\tilde B_\ell+b_3\tilde B_\ell^2\right)
\,.
\ena
To evaluate $(\tilde B-\tilde B_\ell)$ in terms of matter 
densities, 
we expand Eq. (\ref{A2'}) as
\begin{widetext}
\begin{eqnarray*}
&&
\tilde B_\ell C_{\rm m}(\tilde B_\ell)a_g + C_{\rm r}(\tilde B_\ell)
+\Big[\tilde B_\ell C_{\Lambda}'(\tilde B_\ell)
a_g^4 
+
[C_{\rm m}(\tilde B_\ell)+\tilde B_\ell C_{\rm m}'(\tilde B_\ell)]
a_g+ C_{\rm r}'(\tilde B_\ell)
\Big](\tilde B-\tilde B_\ell)
+O((\tilde B-\tilde B_\ell)^2)=0
\,,
\end{eqnarray*}
where we use $C_{\Lambda}(\tilde B_\ell)=0$.
In the limit of $\tilde B\rightarrow \tilde B_\ell$, 
if the universe is expanding, 
dropping the higher-order terms of $a_g^{-1}$
because $a_g$ is increasing, 
we find 
\bea
\tilde B-\tilde B_\ell &\approx 
&
-{1\over  C_{\Lambda}'(\tilde B_\ell)}
\left[{ C_{\rm m}(\tilde B_\ell)
\over a_g^3}+{C_{\rm r}(\tilde B_\ell)
\over \tilde B_\ell a_g^4} 
\right]
+O\left({1\over a_g^6}\right)
\,.
\label{B-Bell}
\ena
Note that $(\tilde B-\tilde B_\ell)\sim O(a_g^{-3})$.
Plugging (\ref{B-Bell}) into (\ref{rho_g_gam_expand}),
we find
\bea
\kappa_g^2 (\rho^{[\gamma]}_g+\rho_g)
&
\approx
&
 \Lambda_g
+\kappa_g^2\rho_g^{\rm (m)}
\left[1-{\Big(1-{\epsilon r_{\rm m}\over \tilde B_\ell}
\Big)\over \Big(1-{2\Lambda_g\over 3 m_{\rm eff}^2}
\Big)\Big(1+{\kappa_f^2\over \tilde B_\ell^2\kappa_g^2}\Big)}\right]
+\kappa_g^2\rho_g^{\rm (r)}
\left[1-{\Big(1-{r_{\rm r}\over \tilde B_\ell^2}
\Big)\over \Big(1-{2\Lambda_g\over 3 m_{\rm eff}^2}
\Big)\Big(1+{\kappa_f^2\over \tilde B_\ell^2\kappa_g^2}\Big)}\right]
+O\left({1\over a_g^6}\right)
\,,~~~
\ena
where 
$\rho_{g,{\rm (m)}}$ and $\rho_{g,{\rm (r)}}$ are energy densities 
of $g$-matter and $g$-radiation, respectively, and 
$r_{\rm r}=c_{f,{\rm r}}/c_{g,{\rm r}}$.
$m_{\rm eff}$ is  the graviton mass in the present background spacetime, which 
is defined by Eq. (\ref{graviton_mass}).
\end{widetext}
Since the energy density of $f$-matter fluids 
in the present limit is approximated by 
\bea
\kappa_f^2\rho_f &=&{c_{f, {\rm m}}\over a_f^3}+
{c_{f, {\rm r}}\over a_f^4}={r_{\rm m}\over \tilde B^3}
 {c_{g, {\rm m}}\over a_g^3}+
{r_{\rm m}\over \tilde B^4}{c_{g, {\rm r}}\over a_g^4}
\nn
&&
\approx {r_{\rm m}\over \tilde B_\ell^3} {c_{g, {\rm m}}\over a_g^3}+
{r_{\rm r}\over \tilde B_\ell^4}{c_{g, {\rm r}}\over a_g^4}
+O(a_g^{-6})
\,,
\ena
replacing $r_{\rm m}$ and $r_{\rm r}$ by $f$-matter fluids, 
we finally obtain a standard form of 
the effective Friedmann equation in the 
present model as
\bea
H_g^2+{k\over a_g^2}={\Lambda_g\over 3}+{\kappa_{\rm eff}^2\over 3}
\left[\rho_g+\rho_{\rm D}\right]
\label{effective_Friedmann_equation}
\ena
where 
\bea
\kappa_{\rm eff}^2&=&
\kappa_g^2
\left[1-{1\over \Big(1-{2\Lambda_g\over 3 m_{\rm eff}^2}
\Big)\Big(1+{\kappa_f^2\over \tilde B_\ell^2\kappa_g^2}\Big)}\right]
\\
\rho_{\rm D}&=&{\kappa_f^2\tilde B_\ell^2\over \kappa_g^2\left[
 \Big(1-{2\Lambda_g\over 3 m_{\rm eff}^2}
\Big)\Big(1+{\kappa_f^2\over \tilde B_\ell^2\kappa_g^2}\Big)-1\right]}
\,\rho_f
\,.
\ena
$\kappa_{\rm eff}^2$ is the effective 
gravitational constant, which is always smaller than the bare 
gravitational constant $\kappa_{g^{2}}$, if the Higuchi bound is satisfied
($m_{\rm eff}^{2}>2\Lambda_{g}/3$).
$\rho_{\rm D}$ is 
regarded as ``dark sector" which origin is another one of twin matter. 
In particular, when dust matter fluids are dominant, 
$\rho_{\rm D}$ is regarded as ``dark matter".

Although the effective Friedmann equation (\ref{effective_Friedmann_equation}) is valid both for
an asymptotic de Sitter universe ($\tilde B_{\ell}=\tilde B_{\rm (dS)}$) and 
for an asymptotic matter dominant universe ($\tilde B_{\ell}=\tilde B_{\rm (M)}=1$),
in what follows, we discuss only the case of $\tilde B_{\rm (dS)}$ to explain 
the present observed universe.

\subsection{Effective Gravitational Constant}
First of all, we discuss about the effective gravitational constant $\kappa_{\rm eff}^{2}$,
which must be positive in order for gravitational force to be attractive.
In Table \ref{grav_cons_mass}, 
we summarize the value of  $\kappa_{\rm eff}^{2}$ as well as $m_{\rm eff}^{2}$
for five models (Models A$\sim$E), where we assume $\kappa_{f}^{2}=\kappa_{g}^{2}$.
From Table \ref{grav_cons_mass}, 
we can reject two models (Models A and C) because those models 
predict a negative gravitational constant.

\begin{table}[H]
\begin{center}
  \begin{tabular}{|c||r|r|}
\hline 
Model
&${\kappa_{\rm eff}^2/ \kappa_g^2}$~~~&
${m_{\rm eff}/m}$
\\ 
\hline 
A&$-0.0880972$&$6.43151$\\  
\hline 
B&$0.976813$&$0.938869$\\  
\hline 
C&$-0.108741$&$3.30578$\\  
\hline 
D&$0.920396$&$0.885782$\\  
\hline 
E&$0.0283764$&$3.99107$\\  
\hline 
 \end{tabular}
\caption{The effective gravitational constant and the effective graviton mass 
in de Sitter background for  Models A $\sim$ E. We assume $\kappa_f^2=\kappa_g^2$.}
\label{grav_cons_mass}
\end{center}
\end{table}

\begin{figure}[h]
\begin{center}
\includegraphics[width=6cm,angle=0,clip]{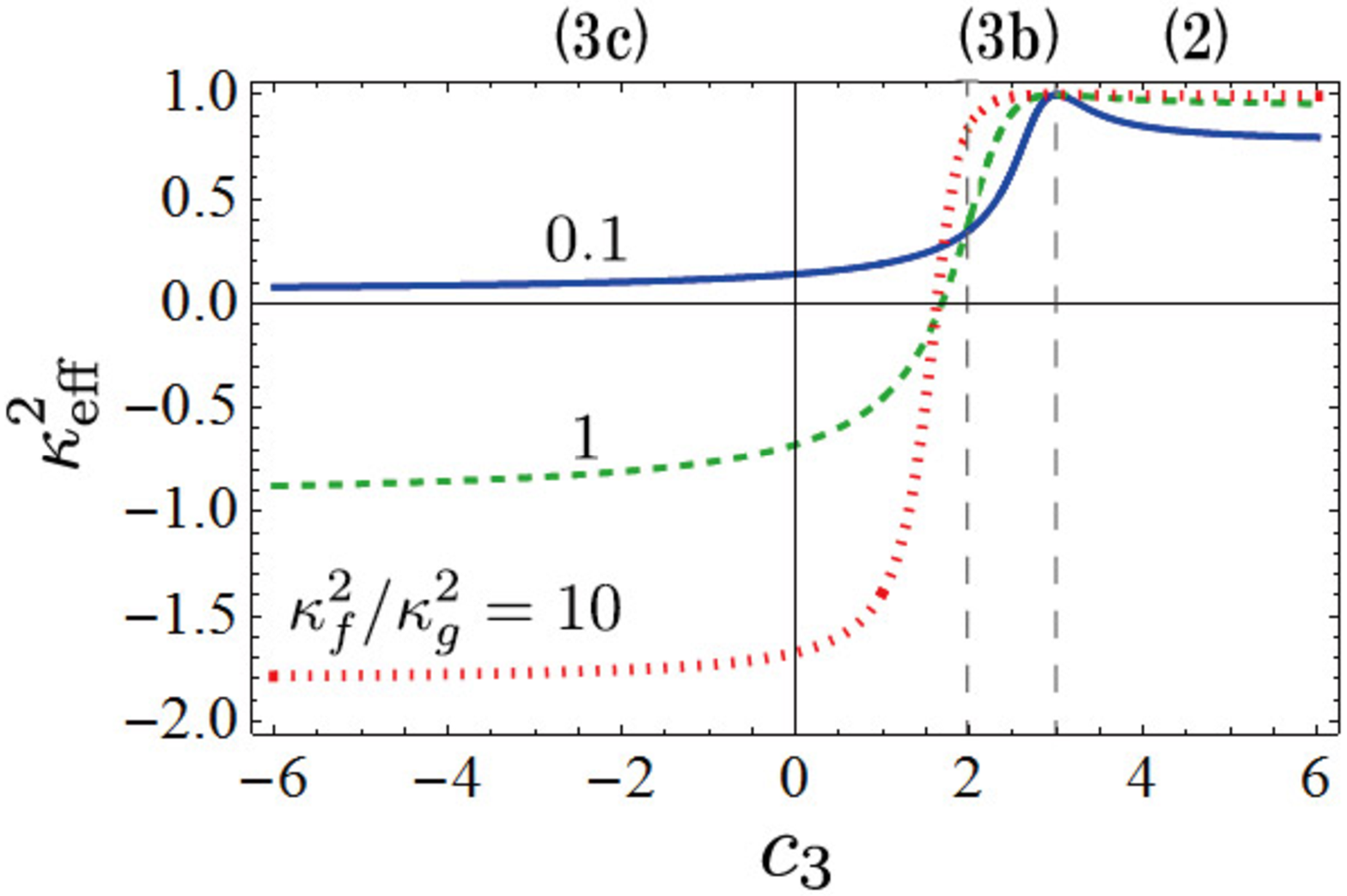}
\\
(a) Case (I)
\\
\includegraphics[width=6cm,angle=0,clip]{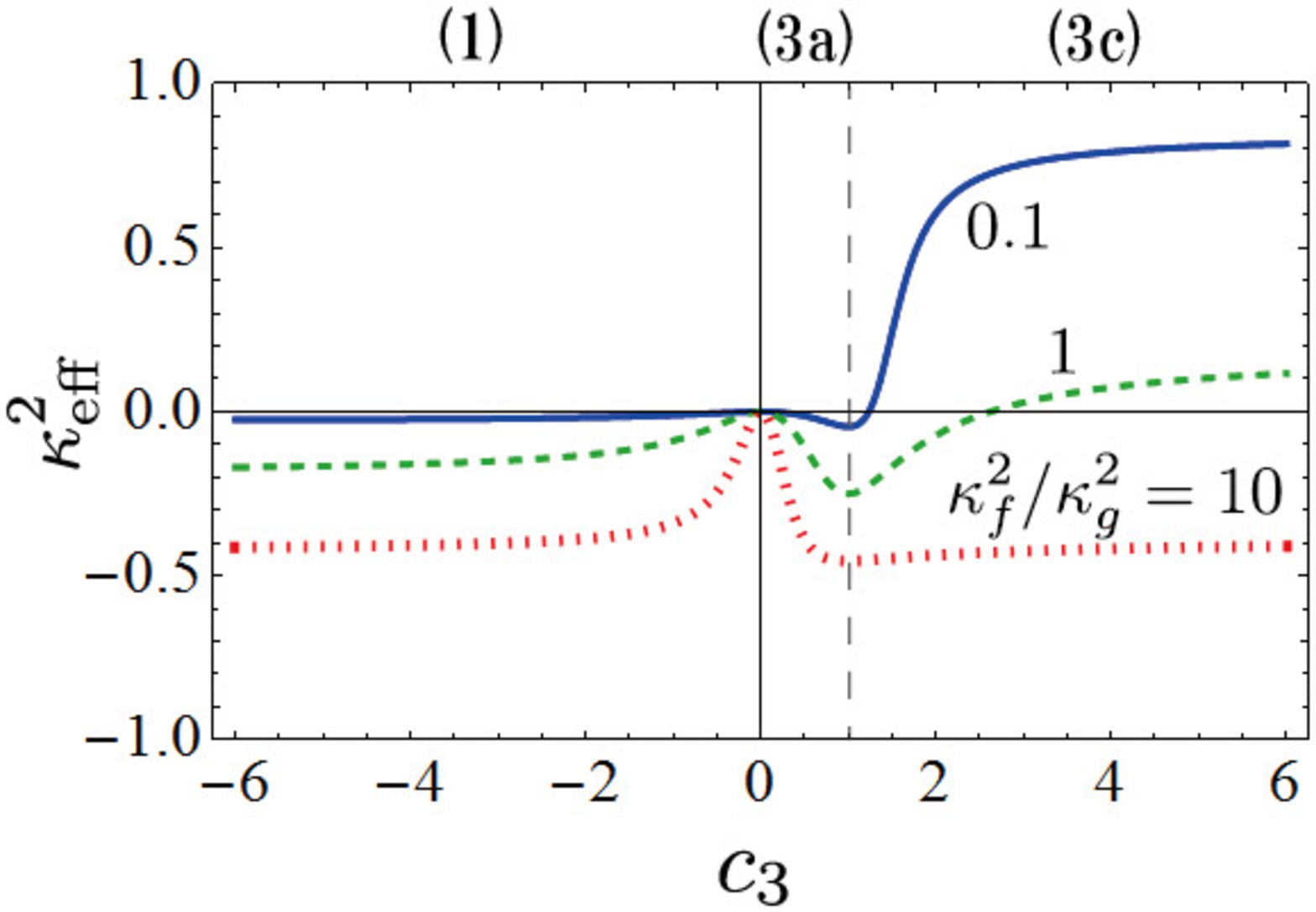}
\\
(b) Case (II)
\\
\caption{The effective gravitational constant $\kappa_{\rm eff}^{2}$ 
 for two one-parameter families of coupling constants
((a) $b_{0}=0$ and (b) $b_{4}=0$). We plot three cases of 
$\kappa_{f}^{2}/\kappa_{g}^{2}=0.1$, $1.0$,
and $10.$.}
\label{eff_grav_const1}
\end{center}
\end{figure}

To see more general cases, we calculate the effective gravitational
constant $\kappa_{\rm eff}^{2}$ for two one-parameter families 
((I) $b_{0}=0$ and (II) $b_{4}=0$),
which we discussed in \S. \ref{cosmic_no_hair}.
Fig. \ref{eff_grav_const1} shows  $\kappa_{\rm eff}^{2}$ with
 respect to $c_{3}$.
We find the constraint on  $c_{3}$ as  $c_{3}> 1.67845$ for the case (I),
 while  $c_{3}> 2.61963$ for the case (II). 

\begin{figure}[h]
\begin{center}
\includegraphics[width=5cm,angle=0,clip]{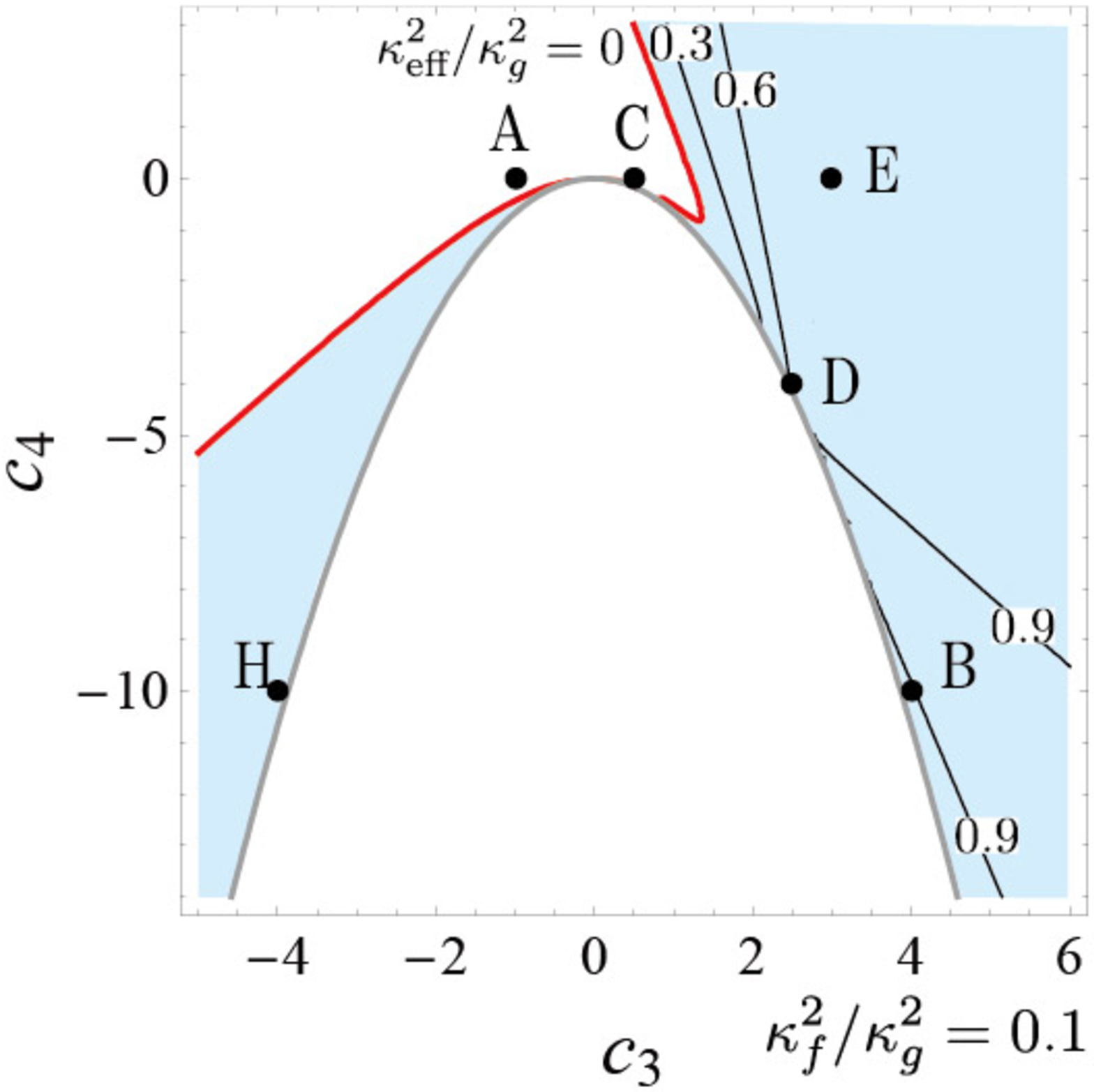}
\\[.5em]
(a) $\kappa_{f}^{2}/\kappa_{g}^{2}=0.1$
\\[.5em]
\includegraphics[width=5cm,angle=0,clip]{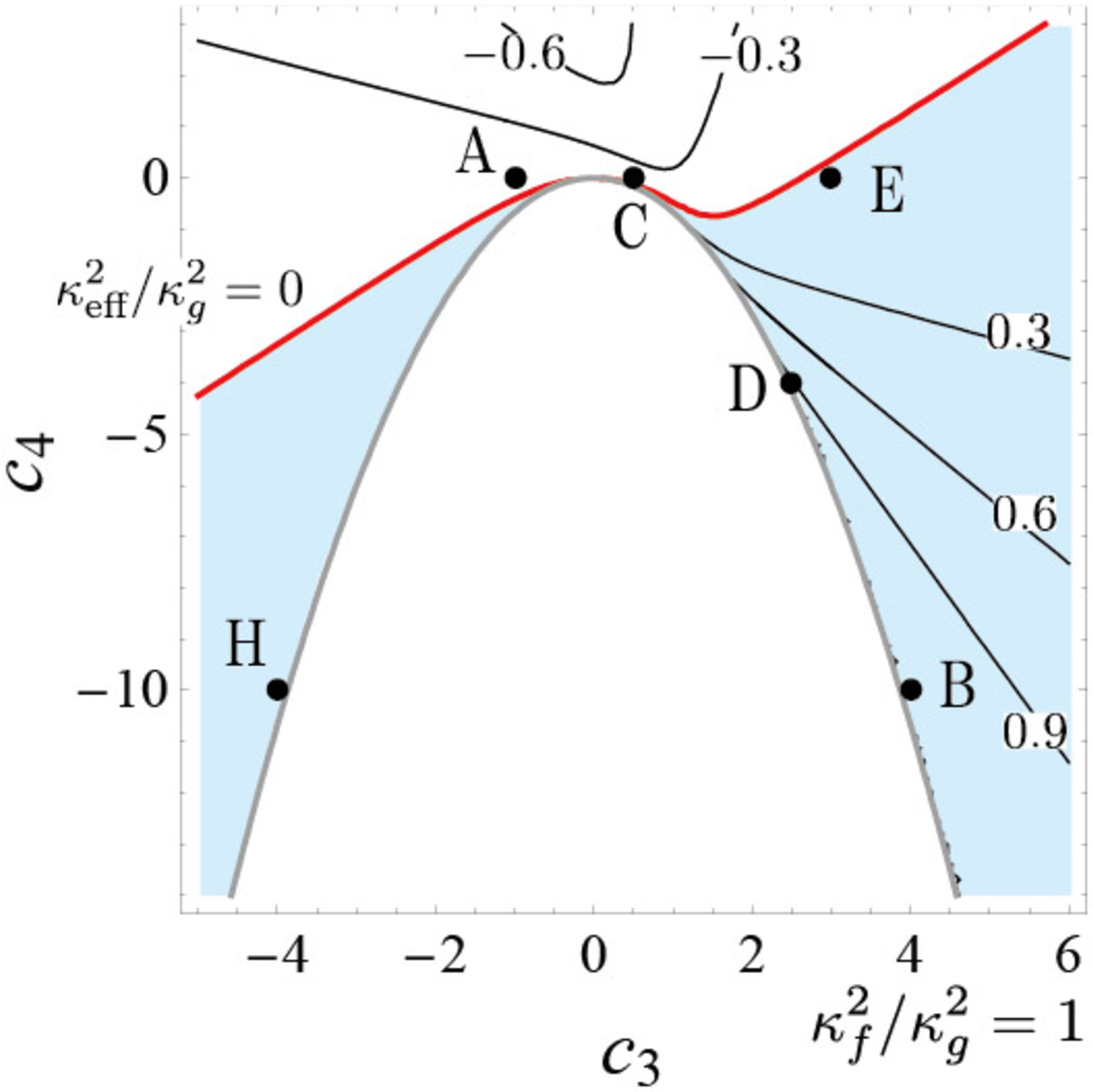}
\\[.5em]
(b) $\kappa_{f}^{2}/\kappa_{g}^{2}=1$
\\[.5em]
\includegraphics[width=5cm,angle=0,clip]{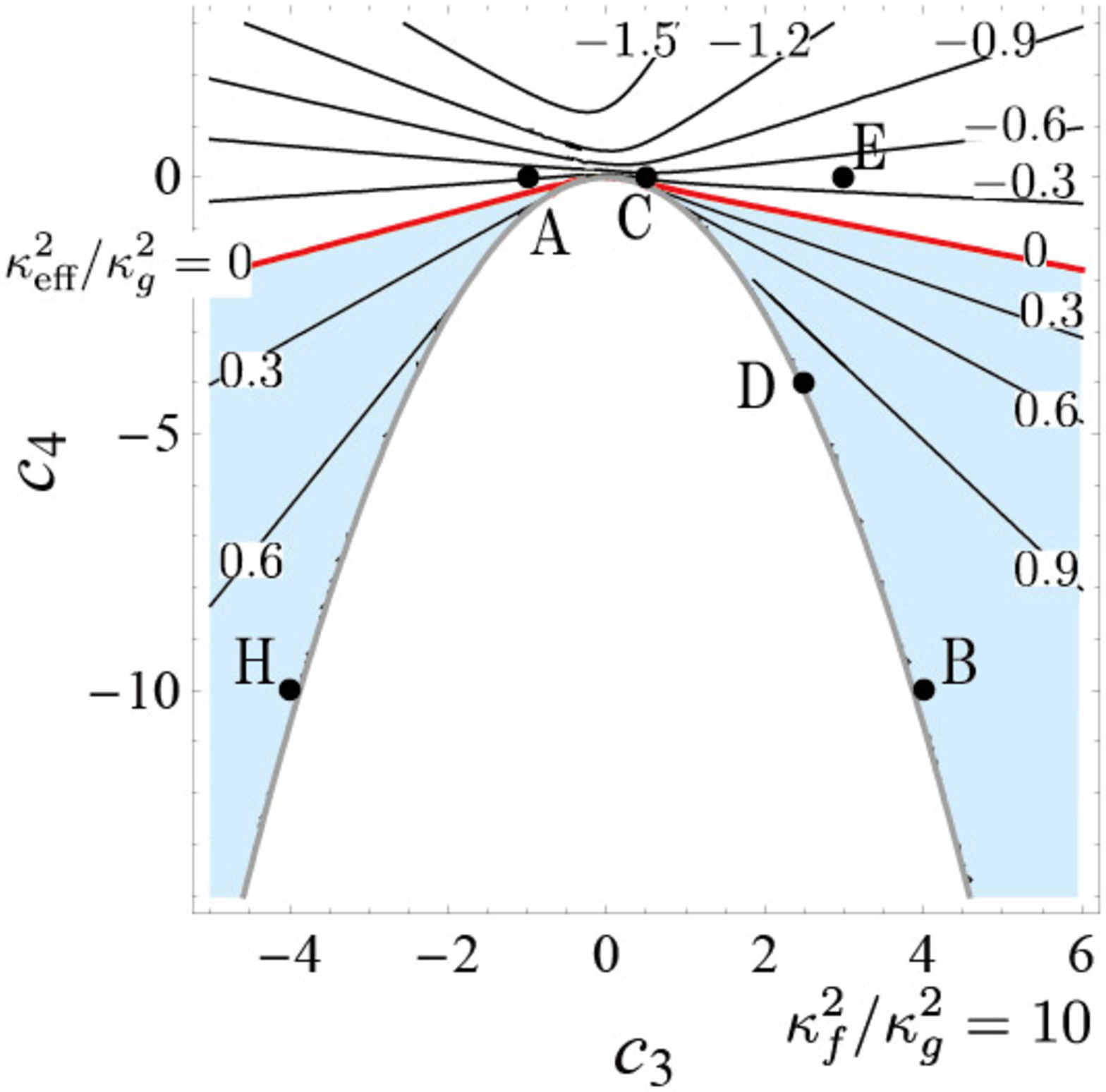}
\\[.5em]
(c) $\kappa_{f}^{2}/\kappa_{g}^{2}=10$
\caption{The contour maps of the 
effective gravitational constant $\kappa_{\rm eff}^{2}$ 
in the $c_{3}$-$c_{4}$ plane for three cases of 
$\kappa_{f}^{2}/\kappa_{g}^{2}=0.1$ (a), $1.0$ (b),
and $10$ (c). The red thick curves denote the contour of 
$\kappa_{\rm eff}^{2}=0$, below which 
(the stripe-shaded light-blue region) 
we find the region of $\kappa_{\rm eff}^{2}>0$,
which is physically required.}
\label{eff_grav_const2}
\end{center}
\end{figure}

For most general parameters, in Fig. \ref{eff_grav_const2}, 
we also show the range of $\kappa_{\rm eff}^{2}>0$ by
the stripe-shaded light-blue region 
in the $c_{3}$-$c_{4}$ plane. 

If we change the ratio of $\kappa_{f}^{2}/\kappa_{g}^{2}$, 
the critical curve for  $c_{3}$ and $c_{4}$ moves.
We show the ranges of the positive gravitational constant 
 for $\kappa_{f}^{2}/\kappa_{g}^{2}=0.1$ and $10$ 
in Fig.\ref{eff_grav_const1} for the cases (I) and (II),
and the contour maps of $\kappa_{\rm eff}^{2}$ in Fig. 
\ref{eff_grav_const2} in the $c_{3}$-$c_{4}$ plane.  
When  $\kappa_{f}^{2}/\kappa_{g}^{2}$ decreases, 
the physically allowed region with a positive effective 
gravitational constant then  increases 
in the range of $c_3>0$, but 
decreases in the range of $c_3<0$, and vice versa.
  
We may have also another constraint on the effective gravitational 
constant.
The effective gravitational constant  $\kappa_{\rm eff}^{2}$ is 
different from 
the bare value $\kappa_{g}^{2}$.  
In particular,  Model E with $\kappa_{f}^{2}=\kappa_{g}^{2}$ gives 
a big discrepancy.
One may wonder whether such a discrepancy is acceptable or not.
Because we know that the difference between the local gravitational 
constant (Newtonian
gravitational constant) and the cosmological one should not be so 
large\cite{Variablity_G}.
If the local gravitational constant is $\kappa_{g}^{2}$ or very 
close to it,
we will find a stronger constraint on the coupling parameters 
($c_{3}$ and $c_{4}$)
as well as the ratio of $\kappa_{f}^{2}/\kappa_{g}^{2}$.
For example, for Model E, if  $\kappa_{f}^{2}/\kappa_{g}^{2}<0.0366125$, 
we find  
$\kappa_{\rm eff}^{2}/\kappa_{g}^{2}>0.9$, which may be consistent 
with observations.
Although we expect that the local gravitational constant is
 close to the bare gravitational constant $\kappa_{g}^{2}$,
 to confirm the above constraint, we have to calculate the local 
gravitational constant 
assuming that the Vainshtein mechanism is working.

\subsection{``Dark Matter''}
Next we discuss the possibility to explain ``dark matter'' component in the Friedmann 
equation by another one of twin matter fluids.
In Table \ref{present_universe}, we show the ratio of ``dark matter'' density $\rho_D$ 
to that of $g$-matter $\rho_{g,{\rm m}}$ for Models B, D and E.
Its value,  of course, depends on the ratio $r_{\rm m}$.
If $\rho_{g,{\rm m}}$ consists only of a baryonic matter,
it gives the ratio of dark matter to a baryonic matter,
which is about 5 from the cosmic pie\cite{Planck}.
So choosing $r_{\rm m}$ appropriately as in Table \ref{present_universe}, 
we find the observed value.
\begin{table}[H]
\begin{center}
  \begin{tabular}{|c||r|}
\hline 
Model
&${\rho_{\rm D}/ \rho_g}$~~~\\ 
\hline 
B&$0.16261\,r_{\rm m}$\\  
\hline 
D&
$0.32278\,r_{\rm m}$\\  
\hline 
E&$44.9613\,r_{\rm m}$\\  
\hline 
 \end{tabular}
\caption{The ratio of the ``dark sector" energy density to the matter
 energy density in $g$-spacetime. $r_{\rm m}$ is the ratio of twin 
matter energy densities.}
\label{present_universe}
\end{center}
\end{table}

However, in order for the de Sitter attractor to be natural, we have the 
constraint on $r_{\rm m}$ as we discussed in \S. \ref{CNHC}.
For example, for Model B, if $r_{\rm m}<r_{\rm m}^{\rm (cr)}=0.41105$, 
the universe approaches to de Sitter spacetime for 
any possible initial value $\tilde B (<\tilde{B}_{\rm (AdS_2 )}=0.59766) $.
This critical value gives $\rho_{D}/\rho_{g}<0.0656841$, which is too small to explain 
the present amount of dark matter. 
To find  $\rho_{D}/\rho_{g}\sim 5$, we need $r_{\rm m}\sim 30$, for which 
a fine-tuning of initial data is required to find de Sitter universe.
Similarly Model D requires a fine-tuning for de Sitter spacetime.
Only Model E gives a model which explains the amount of dark matter as well 
as de Sitter accelerating universe, because any initial value of $\tilde B$ 
leads an de Sitter attractor, assuming $\epsilon=-1$. 
In this case, however, the effective gravitational constant may be too small.
For example,  $\kappa_{\text{eff}}^2/\kappa_g^2=0.0283764$ for 
$\kappa_g^2=\kappa_f^2$. 
Although this value can become close to $\kappa_g^2$ if we choose 
 $\kappa_{f}^{2}/\kappa_{g}^{2}\ll 1$, 
we need a fine-tuning of initial data to find de Sitter universe.
 (see Fig.\ref{kappa} (b).)

\begin{figure}[htbp]
\begin{center}
\includegraphics[width=5.5cm,angle=0,clip]{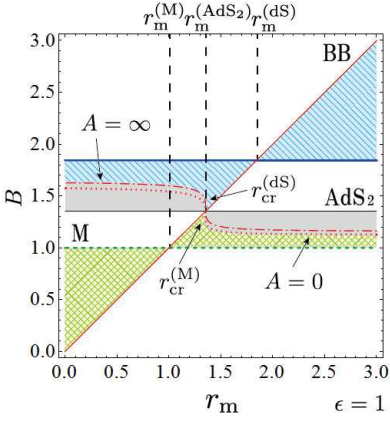}
\caption{The same figure as Fig. \ref{region1} for Model HI
 ($c_3=-4,c_4=-10$ and $\kappa_{f}^{2}/\kappa_{g}^{2}= 60$).
We have 
$r_{\rm m}^{\rm (AdS_2)}=1.35254$, 
$r_{\rm m}^{\rm (dS)}=1.84303$, 
$r_{\rm cr}^{\rm (M)}=1.34987$,
and $r_{\rm cr}^{\rm (dS)}=1.35313$.
}
\label{region1-2}
\end{center}
\end{figure}

More natural model is found if we choose the coupling constants in the 
left-bottom region in the $c_{3}$-$c_{4}$ plane.
One example is Model H with $c_{3}=-4, c_{4}=-10$, which is 
 plotted by the dot H in Fig. \ref{eff_grav_const2}, and which 
data and properties are given  in Table 
\ref{epsilon5} and  in Fig. \ref{region1-2}.

\begin{table}[H]
\begin{center}
  \begin{tabular}{|c|c||c|c|c|r|c|}
\hline 
Model&$(c_3,c_4$)&region&$\epsilon$&$B_\ell$
&~~~~$\Lambda_g$~~~~~~&vacuum \\
\hline 
\hline
\\[-1em]
H& $(-4, -10)$&(1)&$-1$&$-42.9813$&
$-1.22 \times 10^6 m_g^2$&AdS$_1$ 
\\ \cline{4-7} 
&&&1&1&0&M  \\ \cline{4-7} 
&&&$1$&$1.35254$&$-0.193237m_g^2$&AdS$_2$ \\ \cline{4-7} 
&&&$1$&$1.84303$&$0.256594m_g^2$&dS \\ 
\hline 
 \end{tabular}
    \caption{$B_\ell$ and $\Lambda_g$ for Model H.}
\label{epsilon5}
\end{center}
\end{table}

 We present two following examples for the appropriate values of 
 $\kappa_{f}^{2}/\kappa_{g}^{2}$ and the ratio $r_{\rm m}$ : \\
\bea
{\rm Model~HI}~ &:& \kappa_{f}^{2}/\kappa_{g}^{2}= 60
\,,~~
r_{\rm m} = 180
\,,
\nn
{\rm Model~HII}
&:&
\kappa_{f}^{2}/\kappa_{g}^{2}= 1000
\,,~~
r_{\rm m} = 3000
\,.
\ena
We find $\kappa_{\rm eff}^{2}/\kappa_{g}^{2}= 0.946314$ 
and $\rho_{\rm D}/ \rho_g = 5.54065$ 
for Model HI, while  $\kappa_{\rm eff}^{2}/\kappa_{g}^{2}=0.996608$ 
and $\rho_{\rm D}/ \rho_g=5.41619$ for Model HII,
both of which are consistent with observations.

\subsection{$\Lambda$-CDM Model}
\label{dark_radiation}
Although the ratio of dark matter to baryonic matter is constant,
their total amount is time-dependent.
Hence in order to explain the present ratio of each component in 
the cosmic pie (the pie chart of the content of the Universe), we have to 
analyze the evolution of the universe.
In the earlier stage of the universe, that is, when $\tilde B$ is not close 
to $\tilde B_{\rm dS}$, the Friedmann equation is not described 
by the standard form
 (\ref{effective_Friedmann_equation}).
The interaction term can not be written 
only by the linear combination of $\Lambda_g,\rho_g,$ and $\rho_f$ with 
$\kappa_{\text{eff}}^2$.
So we redefine the density of dark sector $\bar \rho_{\text{D}}$ by 
\bea
\kappa_{\rm eff}^{2}\bar \rho_{\text{D}}=
\kappa_{g}^{2}\rho_{g}^{[\gamma]}-\Lambda_{g}-\kappa_{\rm eff}^{2}\rho_{g}
\,,
\ena
which includes higher-order terms of $(\tilde B-\tilde B_{\rm dS})$. 
Note that $\bar \rho_{\text{D}}\rightarrow \rho_{\text{D}}$ as 
$\tilde B \rightarrow \tilde B_{\rm dS}$,
which provides the present amount of dark matter.

Introducing the density parameters, which  are defined by
\begin{eqnarray*}
&&
\Omega_{\Lambda}
=\frac{\Lambda_g}{3H_{g}^2}\,,~~
\Omega_{\text{D}}=\frac{\kappa_{\rm eff}^{2}\bar \rho_{\rm D}}{3H_{g}^2}\,,~~
\nn
&&
\Omega_{\text{m}}=\frac{\kappa_{\rm eff}^{2}\rho_{g,{\rm m}}}{3H_{g}^2}\,,~~
\Omega_{\rm r}=\frac{\kappa_{\rm eff}^{2}\rho_{g,{\rm r}}}{3H_{g}^2}\,,~~
\Omega_{\rm k}=-\frac{k}{H_{g}^2}
\,,
\end{eqnarray*}
we obtain the Friedmann equation for $g$-spacetime  as 
\begin{equation}
\Omega_{\Lambda}+\Omega_{\text{D}}+\Omega_{\text{m}}
+\Omega_{\rm r}+\Omega_{\rm k}=1
\,.
\end{equation}
From the observation, our universe is almost flat and the radiation energy 
is ignorable. Hence we assume that $\Omega_{\rm r}=0$ and $\Omega_{\rm k}=0$.
The present ratio of dark energy (a cosmological constant) is about 70\%, while 
that of the matter density including dark matter is about 30\%
\cite{Planck}, i.e,
\beann 
\Omega_{\Lambda}|_{0}\simeq 0.7
\,,~~
(\Omega_{\text{D}}+\Omega_{\text{m}}) |_{0}\simeq 0.3
\enann
We also know that the baryonic density is given by 
$\Omega_{\rm b}|_{0}\sim 
0.05$\cite{Planck}.

\begin{figure}[h]
\begin{center}
\includegraphics[width=5.5cm,angle=0,clip]{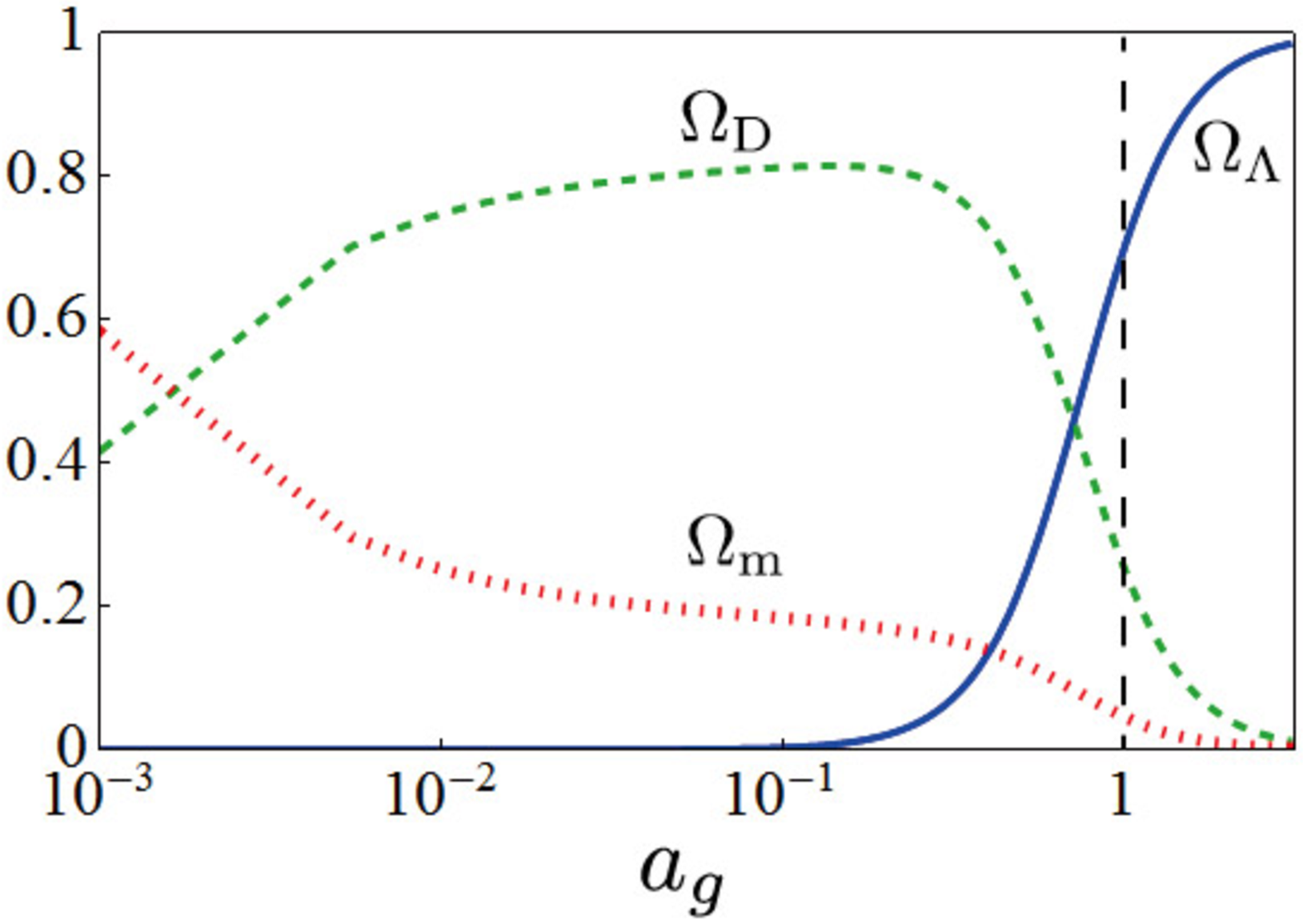}
\\
(a)
\\
\includegraphics[width=5.5cm,angle=0,clip]{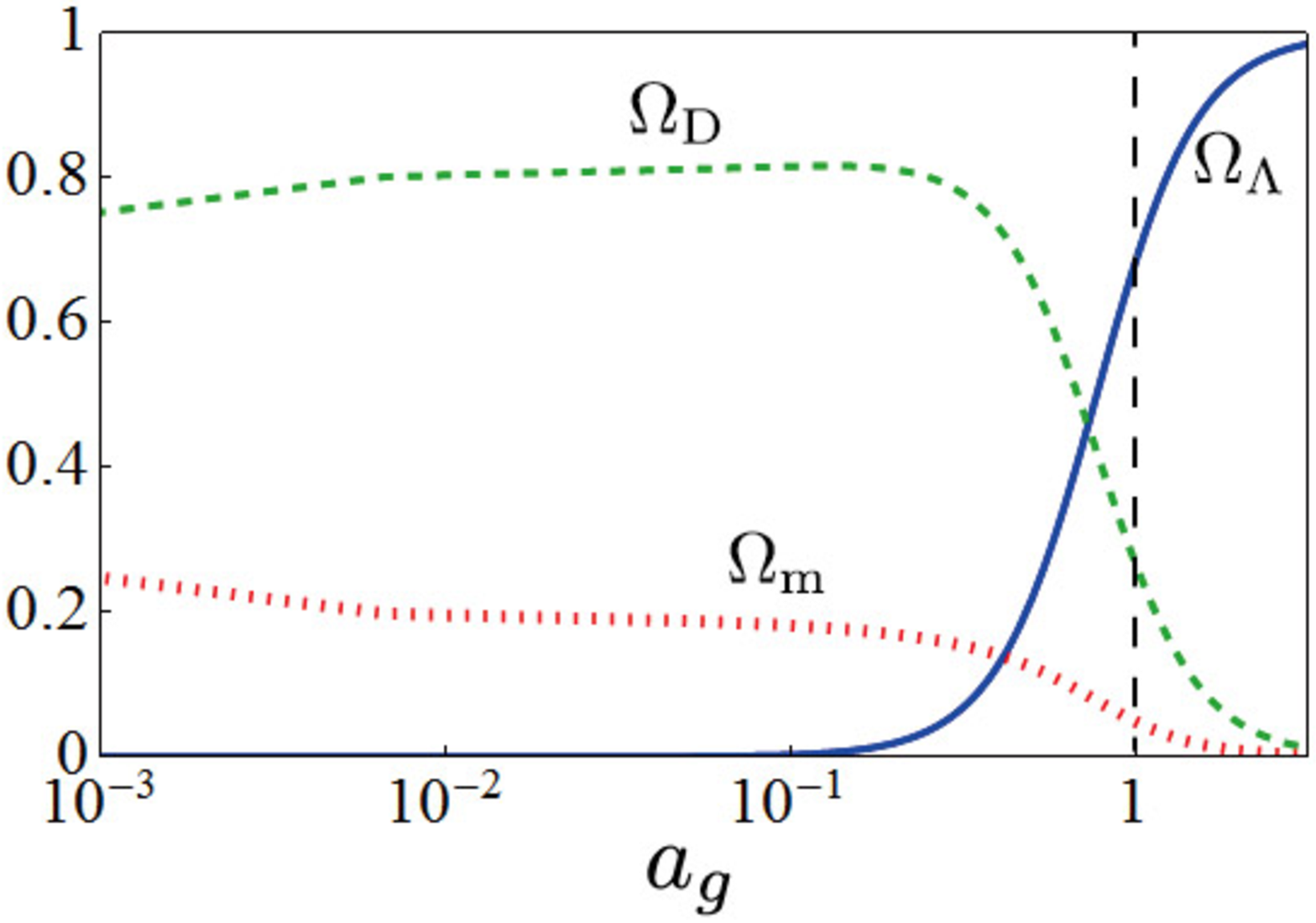}
\\
(b)
\\
\caption{The time evolution of density parameters 
for Model HI ($\kappa_{f}^{2}/\kappa_{g}^{2}= 60
$ and $r_{\text{m}}=180$) and 
for Model HII ($\kappa_{f}^{2}/\kappa_{g}^{2}= 1000
$ and $r_{\text{m}}=3000$). 
 $a_g=1$ is the present time.}
\label{Omega}
\end{center}
\end{figure}

In order to analyze whether our cosmological model is consistent with 
the history of the universe as well as the present observations, 
we show the time evolution of the density parameters.
We choose one successful model with 
the appropriate values of $\kappa_f^2/\kappa_g^2$ and 
$r_{\rm m}$ (Models HI and HII).
In Fig. \ref{Omega}, we show the results for those two models.
The present time, which is shown by the dashed lines in the figures,
 is fixed by the observed value of 
the deceleration parameter 
$q=-\ddot{a}a/\dot{a}^2=-0.527\pm0.026$\cite{Planck}.
We find that 
the present total matter density $(\Omega_{\text{D}}+\Omega_{\text{m}})|_{0}$ 
is about 0.3 and the dark energy $\Omega_{\Lambda}|_{0}$ is about $0.7$,
respectively, 
as shown in Fig. \ref{Omega}.
This result does not depend on the choice of initial value of $\tilde B$.

Since the ratio $\rho_{D}/\rho_{g}\sim 5$ for both models, 
we find $\Omega_{\text{D}}|_{0}\sim 0.25$
and $\Omega_{\text{m}}|_{0}\sim 0.05$, 
which must consist of  baryonic matter because 
$\Omega_{\text{b}}|_{0}\sim 0.05$.
We need not to introduce non-baryonic dark matter in $g$-spacetime.
Another one of twin matter fluids plays a role of dark matter in the 
effective Friedmann equation.
We should, of course, ask whether another one of twin matter fluids can really
 play 
a role of dark matter in the other situations such as the cosmic structure
 formation
or the missing mass in a galactic scale. 
For such a purpose, we have to analyze an inhomogeneous spacetime
(either perturbations or non-linear but Newtonian system).

Since the value of $\tilde{B}$ is finite in any time because of the potential
 form, 
then the interaction energy density $\rho^{[\gamma]}_g(\tilde B)$
  and then the dark matter density $\rho_{\text{D}}$ 
   are also finite. On the other hand,
   the ordinary matter 
  density $\rho_g$ is proportional to $a_{g}^{-3}$ and then 
  it dominates the universe in the early phase. 
  
The equal time when two energy densities become the same 
is after recombination for Model HI while before for Model HII. 
For Model HI, the dark energy density is smaller than the baryonic density 
at recombination.
This fact may show a difficulty of this model in considering the structure
 formation
because the baryon density fluctuation at recombination era is strongly
 constrained by CMB
observation.

We note that the above scenario will be changed 
 if we have large amount of $f$-radiation at present.
Since the dark sector $\rho_{D}$ is dominated by radiation, 
it does not provide ``dark matter'' component.
Hence $\rho_{g}$ term must contain dark matter  as usual scenario.
$\rho_{D}$ gives just a dark radiation, 
which may be strongly constrained\cite{Planck}.


\section{Concluding Remarks}\label{summary}
We have studied the
 dynamics of homogeneous and isotropic FLRW
 spacetime in the ghost-free 
bigravity theory including twin matter sources. 
Assuming the coupling parameters guaranteeing 
the existence of de Sitter space as well as Minkowski spacetime, 
we find two stable attractors for spacetime with ``twin" dust matter 
fields: One is de Sitter accelerating universe and the other is 
matter dominated universe.
We also find the universe with a future singularity 
for some initial data.
However a considerable number of initial data leads to de Sitter 
universe.  Hence, although 
the cosmic no-hair conjecture does not exactly hold, 
the accelerating de Sitter universe is found naturally. 
The $\Lambda$-CDM model is obtained as an attractor.
We also show that the dark matter component 
in the Friedmann equation,
which originates from another twin matter, can be  
about 5 times larger than the baryonic matter, by choosing
the appropriate coupling constants.
For such a model, our matter field consists just of baryons.

One interesting remaining question is whether another twin matter 
can behave really as dark matter.
Dark matter is required not only in the big bang scenario
but also in the cosmological structure formation 
and as dark matter halos existing around galaxies. 
In order to clarify such a question, we have to analyze 
inhomogeneous models, either in a perturbative approach or 
by non-linear analysis. 
The linear perturbation analysis is now in progress.
Another important question is whether the bigravity theory 
will dynamically recover GR with/without a cosmological constant. 
 That is, is a homothetic solution 
an attractor in more general spacetime ?
This question may be related with the above non-linear analysis.
One simple analysis could be performed in a 
spherically symmetric system.
A spherical static spacetime including a black hole has also been studied 
both in the massive gravity and bigravity theories
\cite{BH_massivegravity,MassiveCosmology1,BH_bigravity,ansatze_bigravity,Vainshtein2}.
Although the perturbation analysis 
show the existence of some instability\cite{BH_Perturbation1,BH_Perturbation2},
since the time scale is about the age of the present universe,
we are interested in whether we find a homothetic solution (GR)
in a local dynamical free-fall time scale. 
It is under investigation.

\section*{Acknowledgments}

We would like to thank Takahiro Tanaka for useful discussions.
This work was supported in part by Grants-in-Aid from the 
Scientific Research Fund of the Japan Society for the Promotion of Science 
(No. 25400276). 

\newpage


\appendix
\newpage

\begin{widetext}

\section{The perturbations around the homothetic solution}
\label{perturbation_homothetic}
\end{widetext}
Since the homothetic spacetimes are given by the solutions in GR,
such solutions are important if they are stable.
So we shall discuss the perturbations around such a homothetic solution.
The basic equations are Eqs. (\ref{g-equation}) and (\ref{f-equation}).

The unperturbed spacetimes are assumed to be  
homothetic, i.e.,
\bea
\gB_{\mu\nu}\,,~~{\rm and}~~~  \fB_{\mu\nu}=K^2  \gB_{\mu\nu}
\,,
\ena
which are the solutions of 
\bea
\GB{}^{\mu}_{~\nu}(\gB)&=&-\Lambda_g(K) \delta^\mu_{~\nu}+
\kappa_g^2 \TB{}^{{\rm [m]}\,\mu}_{~~~~~\nu}  
\label{homo_basic_eqs1}
\,,
\\
\cGB{}^{\mu}_{~\nu}(\fB)&=&-\Lambda_f(K) \delta^\mu_{~\nu}+
\kappa_f^2 \cTB{}^{{\rm [m]}\,\mu}_{~~~~~\nu}
 \,,       
\label{homo_basic_eqs2}
\ena
A constant $K$ is determined by a solution of Eq. (\ref{eq_K}), 
and 
\bea
\kappa_f^2 
\cTB{}^{[m]\mu}_{~~~~~\nu}={1\over K^2 }
\kappa_g^2 
\TB{}^{[m]\mu}_{~~~~~\nu}
\,.
\ena

We then consider the following perturbations:
\bea
g_{\mu\nu}&=& \gB_{\mu\nu}+\epsilon h_{\mu\nu}
\,,
\\
f_{\mu\nu}&=&K^2\tilde f_{\mu\nu}=K^2\left(\gB_{\mu\nu}
+\epsilon k_{\mu\nu}\right)
\label{expansion}
\,,
\ena
where $\epsilon \ll 1$.
The sufficies of $k_{\mu\nu}$ 
as well as  $h_{\mu\nu}$ are moved by the background metric
$\gB_{\mu\nu}$.

The energy-momentum tensors of twin matter fluid 
and those from the interaction terms can be expanded as
\bea
\kappa_g^2 T^{[m]\mu}_{~~~~\nu}&=&\kappa_g^2 \left[
\TB{}^{[m]\mu}_{~~~~\nu}+\epsilon \Tf{}^{[m]\,\mu}_{~~~~~\nu}+\cdots\right]
\notag
\\
\kappa_f^2 {\cal T}^{[m]\mu}_{~~~~\nu}&=&\kappa_f^2 \left[
\cTB{}^{[m]\mu}_{~~~~\nu}+ \epsilon \cTf{}^{[m]\, \mu}_{~~~~~\nu}
+\cdots\right]
\,,
\ena
and 
\bea
\kappa_g^2 T^{[\gamma]\mu}_{~~~~\nu}&=&
 -\Lambda_g \delta^\mu_{~\nu}
+\epsilon \kappa_g^2 \Tf{}^{[\gamma]\mu}_{~~~~\nu}
\\
\kappa_f^2 {\cal T}^{[\gamma]\mu}_{~~~~\nu}&=&
-\Lambda_f \delta^\mu_{~\nu}+\epsilon \kappa_f^2 \cTf{}^{[\gamma]\mu}_{~~~~\nu}
\,,
\ena
respectively, 
where
\bea
\kappa_g^2 \Tf{}^{[\gamma]\mu}_{~~~~\nu}
\,&=&\, m_g^2\left[\tauf{}^{\mu}_{~\nu}(h,k)
-\delta \overset{\tiny (0)}{\mathscr{U}}(h,k)\delta^\mu_{~\nu}\right]
\\
\kappa_f^2 \cTf{}^{[\gamma]\mu}_{~~~~\nu}
&=&
 -{m_f^2\over K^4} \left[(h-k) \tauf{}^\mu_{~\nu}
+\tauf{}^{\mu}_{~\nu}(h,k)\right]
\,,
\ena
with 
\bea
\tauf{}^{\mu}_{~\nu}&=&-{1\over 2}\Big[
(b_1K+2b_2K^2+b_3 K^3)(h^\mu_{~\nu}-k^\mu_{~\nu})
\nn
&&
+(b_2K^2+2b_3K^3+b_4 K^4)(h-k)\delta^\mu_{~\nu} \Big]
\nn
\delta \overset{\tiny (0)}{\mathscr{U}}&=&-{1\over 2}
(b_1K+3b_2K^2+3b_3 K^3+b_4 K^4)(h-k)
\,,
\nn
h&=&h^\alpha{}_\alpha\,,~~k\,=\,k^\alpha{}_\alpha
\,.
\nonumber
\ena

\begin{widetext}
The first order perturbation equations are given by:
\bea
\gB{}^{\mu\rho}\Rf{}_{\rho\nu}(\varphi)-\RB{}^{\rho (\mu}\varphi_{\nu)\rho}
&=&
\overset{\tiny (1)}{M}{}^{[\rm m]}{}^{\mu}{}_{\nu}
-{1\over 4}m_{\rm eff}^2
\left[2\varphi^{\mu}{}_{\nu}+\varphi\delta^{\mu}{}_{\nu}\right]
\label{eq_massive}
\,,
\\
\gB{}^{\mu\rho}\Rf{}_{\rho\nu}(\psi)-\RB{}^{\rho (\mu}\psi_{\nu)\rho}
&=&
\overset{\tiny (1)}{\cal M}{}^{[\rm m]}{}^{\mu}{}_{\nu}
\label{eq_massless}
\,,
\ena
where we have introduced new variables 
$\varphi_{\mu\nu}$ and $\psi_{\mu\nu}$ from two metric perturbations
as 
\bea
\varphi_{\mu\nu}:=h_{\mu\nu}-k_{\mu\nu}
\,,~~
\psi_{\mu\nu}:=m_f^2 h_{\mu\nu}+K^2 m_g^2 k_{\mu\nu}
\,,
\ena
and  defined by 
\bea
\overset{\tiny (1)}{M}{}^{[\rm m]}{}^{\mu}{}_{\nu}&=&
\kappa_g^2 \left[\Tf{}^{[\rm m]}{}^{\mu}{}_{\nu}
-{1\over 2}\Tf{}^{[\rm m]}\delta^{\mu}{}_{\nu}\right]-
K^2 \kappa_f^2 \left[
\cTf{}^{[\rm m]}{}^{\mu}{}_{\nu}
-{1\over 2}\cTf{}^{[\rm m]}\delta^{\mu}{}_{\nu}\right]
\notag
\\
\overset{\tiny (1)}{\cal M}{}^{[\rm m]}{}^{\mu}{}_{\nu}&=&
{\kappa_g^2\kappa_f^2
\over \kappa^2}m^2  \left[\left(\Tf{}^{[\rm m]}{}^{\mu}{}_{\nu}
-{1\over 2}\Tf{}^{[\rm m]}\delta^{\mu}{}_{\nu}\right)+
K^4 \left(
\cTf{}^{[\rm m]}{}^{\mu}{}_{\nu}
-{1\over 2}\cTf{}^{[m]}\delta^{\mu}{}_{\nu}\right)\right]
\,.
\ena
$\psi_{\mu\nu}$ and $\varphi_{\mu\nu}$ describe a massless 
and massive modes, respectively.
$m_{\rm eff}$ denotes graviton mass of the massive mode in the homothetic 
background spacetime, which is given by 
\bea
m_{\rm eff}^2&=&\left(m_g^2+{m_f^2\over K^2}\right)
(b_1K+2b_2K^2+b_3 K^3)
\,.
\ena
If the background is the Minkwoski spacetime ($K=1$),
we find $m_{\rm eff}=m$.

$\Rf{}_{\mu\nu}$ is the linear perturbation operator 
of the Ricci tensor,
which is defined for metric perturbation $h_{\mu\nu}$ 
by 
\bea
\Rf_{\mu\nu}(h)={1\over 2}\left[
-\nablaB_\mu \nablaB_\nu h
-\BoxB
h_{\mu\nu}+ \nablaB{}^\alpha(
\nablaB_\nu h_{\alpha\mu}) +\nablaB{}^\alpha
(\nablaB_\mu h_{\alpha\nu}) \right]
\,.
\ena

The Bianchi identity ($\nabla_\mu G^\mu{}_\nu=0$)
gives the conservation of ``graviton" $\gamma^\mu{}_\nu$,
i.e.,
\bea
\nabla_\mu T^{[\gamma]}{}^{\mu}{}_{\nu}=0
\,,
\ena
which perturbation gives 
the constraint on the massive mode $\varphi_{\alpha\beta}$:
\bea
\nablaB{}_\mu \kappa_g^2 \Tf{}^{[\gamma]}{}^{\mu}{}_{\nu}
={m_g^2\over 2}(b_1K+2b_2K^2+b_3K^3)\left[
-\nablaB_\mu \varphi^\mu{}_\nu+\nablaB_\nu \varphi\right]=0
\,.
\ena
Since $m_{\rm eff}^2\neq 0$,
we find 
\bea
\nablaB_\mu \varphi^\mu{}_\nu=\nablaB_\nu \varphi
\label{pert_varphi-Bianchi}
\,.
\ena
Taking a trace of Eq. (\ref{eq_massive}) and using 
Eq. (\ref{pert_varphi-Bianchi}), we find 
\bea
(3 m_{\rm eff}^2-2\Lambda_g)
\varphi
&=&
\kappa_g^2(2\TB{}^{[\rm m]}_{\alpha\beta} \varphi^{\alpha\beta}
-\TB{}^{[\rm m]}\varphi)+
2\overset{\tiny (1)}{M}{}^{[\rm m]}{}^\alpha_{~\alpha}
\,.
\label{pert_varphi-trace}
\ena
Eqs. (\ref{pert_varphi-Bianchi}) and (\ref{pert_varphi-trace}) 
give five constraint equations on $\varphi_{\alpha\beta}$, 
which is consistent with five degrees of freedom for 
the massive gravition mode.
There is no gauge freedom because $\varphi_{\alpha\beta}$ 
is a gauge invariant tensor. 

Using these constraints, we rewrite the above perturbation equations as
\bea
\nablaB{}_\mu \nablaB_\nu\, \varphi 
-\BoxB \varphi_{\mu \nu}
-2\RB{}_{\mu}{}^\alpha{}_{\nu}{}^\beta \varphi_{\alpha\beta}
+m_{\rm eff}^2
\Big(\varphi_{\mu\nu}+{1\over 2}\varphi\gB_{\mu\nu}\Big)
&=&
2\overset{\tiny (1)}{M}{}^{[\rm m]}_{\mu\nu}
\,,
\label{pert_varphi}
\\
-\nablaB{}_\mu \nablaB_\nu \, \psi 
-\BoxB  \psi_{\mu \nu}
+ 2\nablaB{}_{(\nu}\Big[
\nablaB{}^\alpha \, \psi_{\mu)\alpha}\Big] 
-2\RB{}_{\mu}{}^\alpha{}_{\nu}{}^\beta \psi_{\alpha\beta}
&=&
2\overset{\tiny (1)}{\cal M}{}^{[\rm m]}_{\mu \nu}
\,,
\ena
where we have used 
\bea
\nablaB{}^\alpha(
\nablaB_\nu \chi_{\mu\alpha}) 
&=&\nablaB_\nu(
 \nablaB{}^\alpha\chi_{\mu\alpha}) 
+\RB{}^{~~\alpha\beta}_{\mu~~~~\nu}\chi_{\alpha\beta}
+\RB{}^{\rho}_{~\nu}\chi_{\mu\rho}
\,.
\ena

We shall discuss two decoupled modes separately.
\end{widetext}

\subsection{massless mode $\psi_{\mu\nu}$}

Introducing new variable by
\bea
\bar \psi_{\mu \nu}=\psi_{\mu \nu}-{1\over 2}
\psi \gB_{\mu \nu}
\,,
\ena
and imposing the transverse-traceless 
conditions by use of gauge freedom;
\bea
&&
 \nablaB{}^\mu \bar \psi_{\mu\nu}=0
\,,
\nn
&&
\bar \psi=0
\,,
\ena
we find the perturbation equation as
\bea
\BoxB \bar \psi^{({\rm TT})}_{\mu \nu}
+2\RB{}^{~~\alpha~~\beta}_{\mu~~\nu}
\bar \psi^{({\rm TT})}_{\alpha\beta}
=-2\overset{\tiny (1)}{\cal M}{}^{[\rm m]}_{\mu \nu}
\label{perturbation_eq1-2}
\,.
\ena
In the case of a vacuum spacetime, i.e.,
\bea
&&
\RB{}^\mu{}_\nu=\Lambda_g \delta^\mu{}_\nu
\,,~~\RB=4\Lambda_g 
\nn
&&
\overset{\tiny (1)}{\cal M}{}^{[\rm m]}_{\mu \nu}
=0
\label{vacuum}
\,.
\ena
We obtain the perturbation equation as
\bea
\BoxB \bar \psi^{({\rm TT})}_{\mu \nu}
+2\CB{}^{~~\alpha~~\beta}_{\mu~~\nu}
\bar \psi^{({\rm TT})}_{\alpha\beta}
-{2\over 3}\Lambda_g \bar \psi^{({\rm TT})}_{\mu \nu}
=0
\label{perturbation_eq1-3}
\,.
\ena

If the background is de Sitter spacetime (or a spacetime without 
a Weyl tensor),
we find 
\bea
\BoxB \bar \psi^{({\rm TT})}_{\mu \nu}
-{2\over 3}\Lambda_g \bar \psi^{({\rm TT})}_{\mu \nu}
=0
\label{perturbation_eq1-4}
\,.
\ena
Then we can show that de Sitter expanding spacetime is 
stable against this perturbation mode \cite{CNHC3}.

\subsection{masslive mode $\varphi_{\mu\nu}$}
In the case of vacuum state with Eq. (\ref{vacuum}), 
the trace equation (\ref{pert_varphi-trace}) is now
\bea
(3 m_{\rm eff}^2-2\Lambda_g)
\varphi
=0
\label{pert_varphi-trace2}
\,.
\ena
If $3 m_{\rm eff}^2\neq 2\Lambda_g$, we find $\varphi=0$.
The massive mode $\varphi_{\mu\nu}$ must satisfy 
the transverse and traceless conditions
((\ref{pert_varphi-Bianchi}) and (\ref{pert_varphi-trace2})):
\bea 
\nablaB{}^\mu\varphi_{\mu\nu}=0
\,,~~\varphi=0
\,.
\ena
The perturbation equation is now
\bea
\BoxB \varphi_{\mu \nu}
&+&
2\CB{}_{\mu}{}^\alpha{}_{\nu}{}^\beta \varphi_{\alpha\beta}
\nn
&-&\left[{2\over 3}\Lambda_g 
+m_{\rm eff}^2\right]
\varphi_{\mu\nu}
=
0\,,
\label{pert_varphi3}
\ena
We can show that de Sitter expanding spacetime 
(with zero Weyl curvature) is 
stable against the massive perturbation mode too.

The case of $3 m_{\rm eff}^2=2\Lambda_g $ 
is called ``partially massless",
which contains an additional gauge freedom.
There are only four propagation modes\cite{PM}.

\section{Cosmology in partially massless theory}
\label{cosmology_PM}
In this Appendix, we discuss cosmology in partially massless (PM) 
bigravity theory. 
In the PM theory, all coefficients of $C_{\Lambda}(\tilde B)$
 vanish.
Then we have to deal with this case separately. 
However it turns out to be simpler than the general case. 

The coupling constants can be described by only one free parameter $b_2$ 
such that
\begin{eqnarray}
b_1=b_3=0
\,,~~
b_0=3b_2\,{\kappa_f^2}/{\kappa_g^2}
\,,~~
b_4=3b_2\,{\kappa_g^2}/{\kappa_f^2}
\,. 
\label{PM2}
\end{eqnarray} 

By use of  
the relation \eqref{PM2}, we find the Friedmann equation as 
\begin{eqnarray}
H_g^2  +\frac{k}{a_g^2}  &=&
{\kappa^2_g\over 3}\left(\rho^{[\gamma]}_g+\rho_{g} 
\right)
\,,
\label{PMeq}
\end{eqnarray}
where $\rho_{g}$ is the ordinary matter energy density.
The interaction term gives the energy density of ``gravtion",
$\rho^{[\gamma]}_g$, which is given by
\begin{equation}
\rho^{[\gamma]}_g=\frac{3m^2b_2}{\kappa^2}\left(
\tilde B^2+{\kappa_f^2\over \kappa_g^2}\right)
\,.
\end{equation}

Since the coupling constant  $b_2$ always appears with $m^2$,
it can be absorbed into the definition of $m$,  
fixing the constant as $b_2=1$ in the PM bigravity theory.

The algebraic relation  (\ref{A2'})  between $a_g$ and $\tilde B$
 becomes
\begin{eqnarray}
\tilde B C_{\text{m}}(\tilde B) a_g + C_{\rm r}(\tilde B)=0  \,,
\label{PMeq2}
\end{eqnarray}
which gives 
\begin{eqnarray}
&
\tilde B&=
\tilde B_{\pm}
\nn
&:=&
\pm\frac{c_{f,m}a_g+
\sqrt{c_{f,m}^2 a_g^2
+4 c_{f,r}(c_{g,m}a_g+c_{g,r})}}{2(c_{g,m}a_g+c_{g,r})}
\,,~~~~~
\label{PM_B}
\end{eqnarray}
where we have used the positivity conditions of  
$c_{g,{\rm m}}$, $c_{g,{\rm r}}$, $c_{f,{\rm m}}$, 
and $c_{f, {\rm r}}$. 
In order to write down the effective Friedmann equation from 
Eq. (\ref{PMeq}), assuming the universe is expanding, 
we expand Eq. (\ref{PM_B}) as
\bea
\tilde B_{\pm}&=&\pm\Big[{c_{f,{\rm m}}\over c_{g,{\rm m}}} +
\left({c_{f, {\rm r}}\over c_{f,{\rm m}}} - {c_{f,{\rm m}}
 c_{g,{\rm r}}
\over c_{g,{\rm m}}^2}\right){1\over a_g}
\nn
&&
+
\left(-{c_{f, {\rm r}}^2 c_{g,{\rm m}}\over c_{f,{\rm m}}^3} 
+ {c_{f,{\rm m}} c_{g,{\rm r}}^2\over c_{g,{\rm m}}^3}\right) 
{1\over a_g^2}
\nn
&&
+
\left({2 c_{f, {\rm r}}^3 c_{g,{\rm m}}^2\over c_{f,{\rm m}}^5}
 - {c_{f, {\rm r}}^2 c_{g,{\rm r}}\over c_{f,{\rm m}}^3}
 - {c_{f,{\rm m}} c_{g,{\rm r}}^3\over c_{g,{\rm m}}^4}
\right)
{1\over a_g^3}
\nn
&&
+O\left({1\over a_g^4}\right)
\Big]
\ena
as $a_g\rightarrow \infty$.
Using this equation, we find the effective Friedmann equation 
as
\bea
H_g^2 +\frac{k}{a_g^2}  &\approx &
{\Lambda_g\over 3}+{\kappa_{g}^2\over 3}
\Big(\rho_{g,{\rm m}}+\rho_1+\rho_2+\rho_3
\Big) 
\,,~~
\label{eFeq_PM}
\ena
where 
\begin{eqnarray*}
\Lambda_g&=&3m_g^2\left(
r_{\rm m}^2
+{\kappa_f^2\over \kappa_g^2}\right)
\nn
\kappa_{g}^2\rho_1&=&2 \left(r_{\rm r} - r_{\rm m}^2 \right)
\left({c_{g,{\rm r}}\over c_{g,{\rm m}}}\right)
{1\over a_g}
\nn
\kappa_{g}^2\rho_2&=&\left[{\left(-r_{\rm r} + r_{\rm m}^2 \right)
\left(r_{\rm r} + 3 r_{\rm m}^2 \right)
\over r_{\rm m}^2}\right]
\left({c_{g,{\rm r}}\over c_{g,{\rm m}}}\right)^2
{1\over a_g^2}
\nn
\kappa_{g}^2\rho_3&=&2
\left[{\left(r_{\rm r}- r_{\rm m}^2 \right)
\left(r_{\rm r}^2+ r_{\rm r}r_{\rm m}^2+ 2 r_{\rm m}^4 \right)
\over r_{\rm m}^4}\right]
\left({c_{g,{\rm r}}\over c_{g,{\rm m}}}\right)^3
{1\over a_g^3}
\,,
\end{eqnarray*}
with $r_{\rm m}={c_{f,{\rm m}}/c_{g,{\rm m}}}$ 
and $r_{\rm r}={c_{f,{\rm r}}/c_{g,{\rm r}}}$.
$\rho_3$ is an additional matter, 
$\rho_2$ behaves as a correction of the curvature term, 
and $\rho_1$ describes unusual matter with the equation of 
state $P=-{2\over 3}\rho$. 
Since the $\Lambda$-CDM model with 
zero spatial curvature 
describes the present universe 
very well, 
the last two terms ($\rho_1$ and $\rho_2$)
must be very small. 
Such a condition gives a strong constraint on 
both radiation components.
In particular, 
when we can ignore the radiation terms,
we find the $\Lambda$-CDM model.
No additional dust component comes from the $f$-spacetime matter.
Dark matter must be found in $\rho_{g,{\rm m}}$.
Another one of twin matters is not regarded as dark matter. 
The cosmological constant $\Lambda_g$, however in this case,
 depends on the ratio of twin matter fluids $r_{\rm m}$. 
If $\kappa_f\ll \kappa_g$, 
then $\Lambda
\approx 3{m_g^2}r_{\rm m}^2$.
The cosmological constant depends on matter fluids.
It may give us a hint to solve the so-called ``considence problem",
which is a mystery why the amount of dark energy is 
close to that of matter fluid.


\begin{thebibliography}{99}
\bibitem{footnote}
We call the spacetime with $g$-metric a $g$-spacetime, while
one with $f$-metric a $f$-spacetime.
\bibitem{footnote1}
Assuming that the $g$-metric is the FLRW type but 
the  $f$-metric is just spherically symmetric, 
we find two cases: one is both metrics are diagonal as
we assumed here, and the other is $f$-metric is non-diagonal.
The latter case turns out to be both $g$- and $f$-metrics satisfy 
the conventional Einstein equations with a cosmological constant. 
Since we know FLRW universe in GR very well, we do not discuss this case 
here.
\end{thebibliography}
\end{document}